\documentclass[aps, twocolumn,superscriptaddress,floatfix,preprintnumbers]{revtex4-1}
\usepackage{graphicx,amssymb}
\usepackage{amsmath}
\usepackage{float}
\usepackage{amsmath,amssymb}
\usepackage{mathtools}
\usepackage{amssymb}
\usepackage{simplewick}
\usepackage{physics}
\usepackage{color}

\renewcommand{\S}{{\rm S}}
\newcommand{\N}{{\rm N}}

\newcommand{\sign}{\mathop{\text{sign}}\nolimits}

\renewcommand{\vec}[1]{\mathbf{#1}}
\renewcommand{\Im}{\mathop{\rm Im}}
\renewcommand{\Re}{\mathop{\rm Re}}

\begin{document}
\title{Electron cooling by phonons in superconducting proximity structures}
\author{Danilo Nikoli\'c}
\affiliation{Fachbereich Physik, Universit\" at Konstanz, D-78467, Germany}
\author{Denis M. Basko}
\affiliation{Universit\'e Grenoble Alpes and CNRS, LPMMC, 25 rue des Martyrs, 38042 Grenoble, France}
\author{Wolfgang Belzig}
\affiliation{Fachbereich Physik, Universit\" at Konstanz, D-78467, Germany}
\date{\today}

\begin{abstract}
	We investigate the electron-phonon cooling power in disordered electronic systems with a special focus on mesoscopic superconducting proximity structures. Employing the quasiclassical Keldysh Green's function method, we obtain a general expression for the cooling power perturbative in the electron-phonon coupling, but valid for arbitrary electronic systems out of equilibrium. We apply our theory to several disordered electronic systems valid for an arbitrary relation between the thermal phonon wavelength and the electronic mean free path due to impurity scattering. Besides recovering the known results for bulk normal metals and BCS superconductors, we consider two experimentally relevant geometries of superconductor-normal metal proximity contacts. Both structures feature a significantly suppressed cooling power at low temperatures related to the existence of a minigap in the quasiparticle spectrum. This improved isolation low cooling feature in combination with the high tunability makes such structures highly promising candidates for quantum calorimetry.
\end{abstract}

%\pacs{}

\maketitle
\section{Introduction}
In experiments on quantum thermodynamics it is important to understand the ultimate limits of thermal energy transfer in nanoscale systems. A prime candidate for ultra-low temperature detectors of single heat quanta are so-called proximity thermometers, that consist of normal metals in contact with a superconductor offering a great variability by structuring. An important limiting factor in heat control is the unavoidable coupling of electronic systems out of equilibrium to the phonon bath. A general description for proximity thermometers is still missing and we close that gap in this article. 

At low temperatures the electron-phonon coupling plays an important role in description of heat removal from hot electrons~\cite{Giazotto2006}. Besides the theoretical importance, understanding this effect has a practical meaning in quantum calorimetry~\cite{Eisaman2011,Pekola2013,Karimi2020}. Particularly, fluctuations of the electron-phonon cooling power, related to the electron-phonon thermal conductance by a Nyquist-like relation~\cite{Pekola2018,Karimi2020}, provide a fundamental limitation for the minimum portion of energy which can be detected by its electron heating effect. From the theoretical point of view, the problem of electron cooling by phonons is intimately related to that of ultrasound attenuation by electrons, since both problems are concerned with energy exchange between electrons and phonons.

Numerous experiments on normal metals have shown that the power (typically, per unit volume) transferred from hot electrons at temperature $T_\mathrm{e}$ to cold phonons at $T_\mathrm{ph}$ can be written as $Q(T_\mathrm{e},T_\mathrm{ph}) = Q(T_\mathrm{e}) - Q(T_\mathrm{ph})$, where $Q(T)\propto{T}^p$. 
The well-known result $p=5$ for clean normal metals has been proven  experimentally~\cite{Gantmakher1974,Roukes1985,Wellstood1994} in  agreement with theory~\cite{Price1982, Wellstood1994}.
Electron scattering on impurities modifies the power~$p$. Due to the so-called Pippard ineffectiveness condition~\cite{Pippard1955}, disordered metals with fully screened Coulomb interaction have a power $p=6$ at low temperatures, so that the cooling power is weaker than in the clean case~\cite{Schmid1973, Sergeev1986, Sergeev1987, Belitz1987, Yudson2003}, as has been verified experimentally~\cite{Gershenson2001, Savin2004}.
The crossover between the $T^5$ and $T^6$ behaviors occurs at a temperature when the thermal phonon wavelength $\lambda_\mathrm{ph}$ is of the order of the electronic mean free path $\ell$ due to the impurity scattering.

The energy exchange between electrons and phonons has also been studied in bulk BCS superconductors~\cite{Tsuneto1960,Sergeev1986, Shtyk2013, Shtyk2015} and superconducting proximity structures~\cite{Heikkilae2009}. The presence of a gap in the quasiparticle spectrum leads to a significant suppression of the cooling power at low temperatures and makes these systems advantageous for quantum calorimetry applications.
In Ref.~\cite{Heikkilae2009}, the authors studied the influence of the proximity effect on the cooling power by solving the kinetic equation with the electron-phonon collision integral in the clean limit (i.~e., $\ell\gg\lambda_\mathrm{ph}$).
 
In this Article, we calculate the energy current between electrons and phonons, kept at temperatures $T_\mathrm{e}$ and $T_\mathrm{ph}$, respectively, in superconducting proximity structures for an arbitrary relation between $\ell$ and $\lambda_\mathrm{ph}$. We only assume~$\ell$ (i)~to be small compared to the superconducting coherence length and to the typical size of the structure, and (ii)~to be large compared to the Fermi wavelength, so that the proximity effect can be described by the quasiclassical diffusive Usadel equation. Under these conditions, the energy exchange between electrons and phonons is local on the scale~$\ell$. As a result, the spatial dependence of the cooling power is disentangled from its dependence on the phonon momentum. Interstingly, the latter dependence is the same as for a normal metal~\cite{Pippard1955, Tsuneto1960, Schmid1973}.
 
The Article is organized as follows. In Sec.~\ref{sec:theory} we first specify the electron-phonon interaction in the co-moving frame of reference, and present a very general expression for the cooling power in terms of the electronic stress response function. This expression is perturbative in the electron-phonon interaction, but valid for an arbitrary out-of-equilibrium electronic system. Then we show how this stress response function can be found in a proximitized superconducting structure using the quasiclassical Keldysh Green's formalism. The central result of this chapter is the cooling power expression mentioned above. In Sec.~\ref{sec:results} we illustrate this approach by applying it to several electronic structures. Besides recovering the known results for a bulk normal metal and a bulk BCS superconductor, we consider two geometries of mesoscopic superconductor-normal metal proximity structures: a normal metal tunnel-coupled to a bulk superconductor and a bilayer of a normal metal in contact with a superconductor. We find a strong suppression of the cooling power at low temperatures that is related to the formation of a minigap in the spectrum. Finally, in Sec.~\ref{sec:conclusion} we summarize our work and give concluding remarks. 
\textsl{}
\section{General framework}
\label{sec:theory}

\subsection{Electron-phonon interaction and cooling power}
\label{subsec:e-ph}

Since we are going to describe several structures with normal and superconducting parts, we do not specify the electronic Hamiltonian here. We assume the electrons to be in the diffusive limit because of impurity scattering, and the Coulomb interaction is assumed to be very strong. The electrons are described by the usual fermionic field operators $\hat{\psi}^\dagger(\vec{r})$ and $\hat{\psi}(\vec{r})$. We omit the spin indices for compactness (the spin multiplicity will give an additional factor of 2 in the final result). Eventually, we will only need the electronic quasiclassical Keldysh Green's functions, built from these electronic operators, and satisfying the Eilenberger equation in the presence of impurities. 

The acoustic phonons are described via the lattice displacement field $\hat{\vec{u}}(\vec{r})$, giving the displacement of an atom initially located at the point~$\vec{r}$. In the standard quantization procedure for  lattice vibrations, the displacement operator takes the following form:
\begin{equation}
\label{eqn:displacement}
\hat{\vec{u}}(\vec{r})=\sum_{\vec{q}\lambda}\vec{e}^{\vec{q}\lambda}\sqrt{\frac{1}{2\rho_0L^3\omega_{\vec{q}\lambda}}}\left(\hat{b}_{\vec{q}\lambda} + \hat{b}^\dagger_{\vec{-q}\lambda}\right)e^{i\vec{q}\vec{r}},
\end{equation}
where $\hat{b}_{\vec{q}\lambda}~(\hat{b}^\dagger_{\vec{q}\lambda})$ is the annihilation (creation) operator of a phonon with momentum $\vec{q}$ and polarization characterized by a unit polarization vector $\vec{e}^{\vec{q}\lambda}$, $\omega_{\vec{q}\lambda}$ is the phonon frequency which in general depends on the momentum and polarization, $\rho_0$ is the mass density of the material and $L^3$ is the sample volume. For the polarization $\lambda=l,t1,t2$ we assume a decomposition in one longitudinal and two transverse modes in a standard manner: the longitudinal $\vec{e}^{\vec{q},l}=\vec{q}/q$, while two transverse vectors $\vec{e}^{\vec{q},t1},\vec{e}^{\vec{q},t2}$ are chosen so that $\vec{e}^{\vec{q},t1}\cdot\vec{q}=\vec{e}^{\vec{q},t2}\cdot\vec{q}=0$ and $\vec{e}^{\vec{q},t1}\cdot \vec{e}^{\vec{q},t2}=0$. The dispersion relation is assumed to be the usual relation for acoustic phonons, $\omega_{\vec{q},l} = c_lq$, $\omega_{\vec{q},t1} = \omega_{\vec{q},t2} = c_tq$, where $c_{l/t}$ is the longitudinal/transverse speed of sound in the material, respectively. Due to isotropy, we assume the two transverse modes to have the same velocity, i.e., $c_{t1}=c_{t2}=c_t$. The lattice Hamiltonian is
\begin{equation}
\hat{H}_\mathrm{ph}=\sum_{\vec{q}\lambda}\omega_{\vec{q}\lambda}\,\hat{b}^\dagger_{\vec{q}\lambda} \hat{b}_{\vec{q}\lambda}.
\end{equation}

As discussed in Refs.~\cite{Tsuneto1960, Schmid1973, Sergeev1986, Shtyk2013, Shtyk2015}, the electron-phonon interaction in disordered systems is most conveniently described in a co-moving reference frame, i.~e., attached to the oscillating ions of the crystal lattice, since the impurities oscillate together with the lattice. The electron-phonon interaction Hamiltonian is assumed to have the form
\begin{equation}
\hat{H}_{\rm e-ph} = \int d\vec{r}
\,\hat{\sigma}_{ij}(\vec{r})\,\hat{u}_{ij}(\vec{r}),
\end{equation}
where $\hat{\sigma}_{ij}$ and $\hat{u}_{ij}$ are the stress and strain tensors, respectively,
\begin{eqnarray}
&&\hat{\sigma}_{ij}(\vec{r}) =\frac{1}{4m}\! \left(\frac{\partial}{\partial r_i} - \frac{\partial}{\partial r'_i}\right) \!
\left(\frac{\partial}{\partial r_j} - \frac{\partial}{\partial r'_j}\right)
\hat{\psi}^\dagger(\vec{r})\,\hat{\psi}(\vec{r'})\bigg|_{\vec{r}=\vec{r'}}
\nonumber\\
&&\qquad\qquad{}+\frac{\delta_{ij}}3\,\frac{p_F^2}m\,\hat{\psi}^\dagger(\vec{r})\,\hat{\psi}(\vec{r}),
\label{eqn:stress_tensor}
\\
\label{eqn:strain_tensor}
&&\hat{u}_{ij}(\vec{r})=\frac{1}{2}\left(\frac{\partial \hat{u}_i}{\partial r_j} + \frac{\partial \hat{u}_j}{\partial r_i}\right),
\end{eqnarray}
and the summation over the repeated Cartesian indices $i,j=x,y,z$ is implied. 
The Coulomb interaction is assumed to be very strong, so that electronic charge density fluctuations are completely suppressed. Formally this is described by dressing the electron-phonon vertex by the Coulomb interaction in the random phase approximation~\cite{Schmid1973, Sergeev1986, Shtyk2013, Shtyk2015} and results in the subtraction from the first term in $\hat{\sigma}_{ij}(\vec{r})$ of its projection on the electron density. Here, $p_F$~is the electron Fermi momentum defined via the average of $p^2$ over the Fermi surface, $p_F=(\langle{p}^2\rangle_F)^{1/2}$, and $m$~the free electron mass.

With the electron-phonon interaction at hand we can define the operator for the total energy current flowing into the phonons due to the electron-phonon interaction in the whole sample:
\begin{equation}
\label{eqn:heat_flux_operator}
%\begin{split}
\dot{\hat{H}}_{\rm ph} = i[\hat{H}_{\rm e-ph}, \hat{H}_{\rm ph}] = i\int d\vec{r} \,[\hat{u}_{ij}(\vec{r}), \hat{H}_{\rm ph}]\,\hat{\sigma}_{ij}(\vec{r}).
%\\
%&=-\sum_{ij}\int d\vec{r}\frac{\partial \hat{u}_{ij}(\vec{r})}{\partial t}\hat{\sigma}_{ij}(\vec{r}) = \sum_{ij}\int d\vec{r}\hat{v}_{ij}(\vec{r})\hat{\sigma}_{ij}(\vec{r}),
%\end{split}
\end{equation}
%where $\hat{v}_{ij} = -\partial \hat{u}_{ij}/\partial t$.
We are interested in the cooling power when the phonons are in thermal equilibrium at temperature $T_\mathrm{ph}$. Since the energy current~(\ref{eqn:heat_flux_operator}) is linear in the phonon operators, it has zero average over any density matrix of the direct product form, $\hat\rho_\mathrm{e}\otimes{e}^{-\hat{H}_\mathrm{ph}/T_\mathrm{ph}}$, with an arbitrary electronic density matrix~$\hat\rho_\mathrm{e}$. To obtain a non-zero value to leading order in the electron-phonon coupling, one has to perturb such a phonon state to the first order in $\hat{H}_{\rm e-ph}$. This amounts to calculating the linear response of $d\hat{H}_{\rm ph}/dt$ to the perturbation $\hat{H}_{\rm e-ph}$, which can be done using the Kubo formula:
\begin{equation}
\label{eqn:Kubo}
P = \left\langle \frac{d\hat{H}_{\rm ph}}{dt} \right\rangle  = -i \int\limits_{-\infty}^t dt' \,\langle [\dot{\hat{H}}_{\rm ph} (t), \hat{H}_{\rm e-ph} (t')] \rangle_0, 
\end{equation}
where all the operators are represented in the interaction picture and the average is taken over the non-interacting density matrix $\hat\rho_\mathrm{e}\otimes{e}^{-\hat{H}_\mathrm{ph}/T_\mathrm{ph}}$. 

Since we are dealing with a non-equilibrium situation, it is natural to use the Keldysh Green's function formalism. 
Expanding the commutator in the Kubo formula and identifying various Keldysh Green's function components \cite{Keldysh1964} (see Appendix~\ref{app:Keldysh}), we end up with the following expression for the cooling power:
\begin{equation}
\label{eqn:Q_ph_K}
\begin{split}
P = &
%%%%%%%%%%%
 \frac{1}{4}\int d\vec{r}\, d\vec{r'} \int\limits_{-\infty}^{\infty} \frac{d\omega}{2\pi}\, \omega\,\\
&{}\times \bigg\{D^K_{ijkl}(\vec{r},\vec{r'},\omega)\left[\Pi^R_{klij}(\vec{r'},\vec{r},\omega)-\Pi^A_{klij}(\vec{r'},\vec{r},\omega)\right]
\\ &{} -
\left[D^R_{ijkl}(\vec{r},\vec{r'},\omega)-D^A_{ijkl}(\vec{r},\vec{r'},\omega)\right]\Pi^K_{klij}(\vec{r'},\vec{r},\omega) \bigg\},
\end{split}
\end{equation}
%
%\begin{widetext}
%\begin{equation}
%\label{eqn:Q_ph_K}
%\begin{split}
%Q_{\rm ph} =
%%%%%%%%%%%%
% \frac{1}{4}\int d\vec{r}\, d\vec{r'} \int\limits_{-\infty}^{\infty} \frac{d\omega}{2\pi}\, \omega\,\bigg\{D^K_{ijkl}(\vec{r},\vec{r'},\omega)&\left[\Pi^R_{klij}(\vec{r'},\vec{r},\omega)-\Pi^A_{klij}(\vec{r'},\vec{r},\omega)\right]
%\\ -&
%\left[D^R_{ijkl}(\vec{r},\vec{r'},\omega)-D^A_{ijkl}(\vec{r},\vec{r'},\omega)\right]\Pi^K_{klij}(\vec{r'},\vec{r},\omega) \bigg\},
%\end{split}
%\end{equation}
%\end{widetext}
where $\Pi^{R,A,K}_{ijkl}(\vec{r},\vec{r'},\omega)$ and $D^{R,A,K}_{ijkl}(\vec{r},\vec{r'},\omega)$ are the retarded, advanced and Keldysh components of Green's functions built from the bosonic operators (\ref{eqn:stress_tensor}) and (\ref{eqn:strain_tensor}), respectively. For phonons in thermal equilibrium at temperature $T_\mathrm{ph}$,
\begin{equation}
D^K_{ijkl}(\omega) =\left[D^R_{ijkl}(\omega) - D^A_{ijkl}(\omega)\right]\coth\frac{\omega}{2T_\mathrm{ph}}.
\end{equation} 
%Plugging the latter formula into Eq.~(\ref{eqn:Q_ph_K}) we arrive at
%\begin{eqnarray}
%\label{eqn:Q_ph_RA}
%%\begin{split}
%Q_{\rm ph} =&& \frac{1}{4}\sum_{ijkl}\int d\vec{r}\int d\vec{r'}\int\limits_{-\infty}^{\infty} \frac{d\omega}{2\pi} \omega\bigg\{D^R_{ijkl}(\vec{r},\vec{r'},\omega)-\nonumber
%\\
%&&D^A_{ijkl}(\vec{r},\vec{r'},\omega)\bigg\}
%\bigg\{\left[\Pi^R_{ijkl}(\vec{r'},\vec{r},\omega)-\Pi^A_{ijkl}(\vec{r'},\vec{r},\omega)\right]\times\nonumber
%\\
%&&\times\coth{\left(\frac{\omega}{2T_\mathrm{ph}}\right)}-\Pi^K_{ijkl}(\vec{r'},\vec{r},\omega)\bigg\}. 
%%\end{split}
%\end{eqnarray}
Being mainly interested in various proximity structures, we can make further assumptions  about the spatial dependence. Assuming the phonons to be unaffected by the proximity effect, we take them spatially homogeneous, $D_{ijkl}(\vec{r},\vec{r'},\omega) = D_{ijkl}(\vec{r}-\vec{r'},\omega)$. On the other hand, the situation for the electronic polarization operator, $\Pi_{ijkl}(\vec{r},\vec{r'},\omega)$ can be rather complicated, due to the proximity effect. Still, we assume the electrons to be in the quasiclassical regime, so we pass to the Wigner (mixed) representation where the spatial dependence of $\Pi_{ijkl}(\vec{r},\vec{r'},\omega)$ is decomposed into the center of mass, $\vec{R}=(\vec{r}+\vec{r'})/2$, and the relative coordinate component, $\vec{x}=\vec{r}-\vec{r'}$. Fourier transform is, therefore, performed in the following way
\begin{align}
&D_{ijkl}(\vec{r},\vec{r'},\omega)=\int \frac{d\vec{q}}{(2\pi)^3}\,D_{ijkl}(\vec{q},\omega)\,e^{i\vec{q}(\vec{r}-\vec{r}')},
\\
&\Pi_{ijkl}\left(\vec{R}+\frac{\vec{x}}{2},\vec{R}-\frac{\vec{x}}{2},\omega\right)=\int \frac{d\vec{q}}{(2\pi)^3}\,\Pi_{ijkl}(\vec{R},\vec{q},\omega)\,e^{i\vec{q}\vec{x}}.
\end{align}
In leading order of the quasiclassical approximation, the spatial convolution becomes a simple product in the Wigner representation. Then the total cooling power can be written as a volume integral, $P=\int{d}\vec{R}\,Q(\vec{R})$, where the position-dependent power per volume is given by
\begin{widetext}
%\begin{equation}
%\label{eqn:Q_ph_q}
%\begin{split}
%P = \frac{1}{4}&\int d\vec{R}\int \frac{d\vec{q}}{(2\pi)^3}\int\limits_{-\infty}^{\infty} \frac{d\omega}{2\pi}\, \omega\left[D^R_{ijkl}(\vec{q},\omega)-D^A_{ijkl}(\vec{q},\omega)\right]
%\times
%\\
%&{}\times
%\left\{\left[\Pi^R_{klij}(\vec{R},\vec{q},\omega)-\Pi^A_{klij}(\vec{R},\vec{q},\omega)\right]
%\coth\frac{\omega}{2T_\mathrm{ph}}-\Pi^K_{klij}(\vec{R},\vec{q},\omega)\right\}.
%\end{split}
%\end{equation}
\begin{equation}\label{eqn:Q_ph_q}
Q(\vec{R}) = \int \frac{d\vec{q}}{(2\pi)^3}\int\limits_{-\infty}^{\infty} \frac{d\omega}{2\pi}\, \frac\omega{4}\left[D^R_{ijkl}(\vec{q},\omega)-D^A_{ijkl}(\vec{q},\omega)\right]
\left\{\left[\Pi^R_{klij}(\vec{R},\vec{q},\omega)-\Pi^A_{klij}(\vec{R},\vec{q},\omega)\right]
\coth\frac{\omega}{2T_\mathrm{ph}}-\Pi^K_{klij}(\vec{R},\vec{q},\omega)\right\}.
\end{equation}
\end{widetext}
This formula is very general and applicable to an arbitrary electronic system out of equilibrium. 
%Since we already mentioned that in CFR we deal with free phonons the treatment of the phononic spectral function, $D^R-D^A$, is rather simple.  
Combining Eqs.~(\ref{eqn:strain_tensor}) and (\ref{eqn:displacement}), we obtain the phonon spectral function explicitly:
%%\begin{widetext}
%\begin{equation}
%\label{eqn:spectral_D}
%\begin{split}
%D^R_{ijmn}(\vec{q},\omega)-D^A_{ijmn}(\vec{q},\omega)= &
%\frac{q_iq_jq_mq_n}{q^4} \left(\frac{q\pi i}{\rho_0c_l}\right)\left[\delta(\omega+c_lq) -\delta(\omega-c_lq)\right] +
%%%%%%%%%%%
%\\
%%%%%%%%%%%
%& {}+\sum_{\lambda=t1,t2}\frac{(e_\lambda^iq_j+e_\lambda^jq_i)(e_\lambda^mq_n+e_\lambda^nq_m)}{4q^2}\left(\frac{q\pi i}{\rho_0c_t}\right)\left[\delta(\omega+c_tq) -\delta(\omega-c_tq)\right],
%\end{split}
%\end{equation}
%\end{widetext}
%where $e_\lambda^i$ denotes $i$-th Cartesian component of $\vec{e}_\lambda$.
\begin{subequations}\label{eqn:spectral_D}
\begin{align}
&D^R_{ijkl}(\vec{q},\omega)-D^A_{ijkl}(\vec{q},\omega)=\sum_{\lambda}
\mathcal{T}^{\vec{q}\lambda}_{ij}\,\mathcal{T}^{\vec{q}\lambda}_{kl}\,
D_\lambda^{R-A}(\vec{q},\omega),\\
&D_\lambda^{R-A}(\vec{q},\omega)=i\pi\,\frac{q}{\rho_0c_\lambda}\left[\delta(\omega+c_\lambda{q}) -\delta(\omega-c_\lambda{q})\right],
\label{eqn:Dlambda}\\
&\mathcal{T}^{\vec{q},l}_{ij}\equiv\frac{q_iq_j}{q^2},\quad
\mathcal{T}^{\vec{q},t\kappa}_{ij}\equiv\frac{e^{\vec{q},t\kappa}_iq_j+e^{\vec{q},t\kappa}_jq_i}{2q}\quad(\kappa=1,2).
\end{align}\end{subequations}

\subsection{Electronic polarization operator}
\label{subsec:polarization_operator}

Electrons in the proximitized superconductor are described by the quasiclassical Green's function $\check{g}(t,t';\vec{r},\vec{n})$ with $\vec{n}$ and $\vec{r}$ being, respectively, the unit vector which indicates the direction of momentum and the center of mass coordinate \cite{Larkin1969, Belzig1999}. The Green's function has $2\times2$ matrix structure in Keldysh space:
\begin{equation}\label{eq:checkg=}
\check{g}
=\left(\begin{array}{cc} \hat{g}^R & \hat{g}^K \\ \hat{g}^Z & \hat{g}^A \end{array}\right),
%=\left(\begin{array}{cccc}
%g^R & f^R & g^K & f^K \\ \bar{f}^R & -g^R & \bar{f}^K & -g^K \\
%0 & 0 & g^A & f^A \\ 0 & 0 & \bar{f}^A & -g^A \\
%\end{array}\right),
\end{equation}
where each component is itself a $2\times2$ matrix in the Gor'kov-Nambu space~\cite{Belzig1999}.
The Green's function is subject to the constraint
\begin{equation}
\label{eqn:gg=1}
\int\check{g}(t,t'';\vec{r},\vec{n})\,\check{g}(t'',t';\vec{r},\vec{n})\,dt''
=\check{1}_{4\times4}\,\delta(t-t'),
\end{equation}
and satisfying the Eilenberger equation~\cite{Eilenberger1968}:
\begin{equation}
\label{eqn:Eilenberger}
\left[\hat\tau_3\,\partial_t+\hat\tau_1\Delta+i\hat\tau_0\check{V}+\frac{\langle\check{g}\rangle_\vec{n}}{2\tau},\check{g}\right]+v_F\vec{n}\cdot\vec\nabla\check{g}=0,
\end{equation}
where $\hat\tau_{i}$ are the Pauli matrices in the Gor'kov-Nambu space and $\langle\ldots\rangle_\vec{n}$ denotes the average over the directions~$\vec{n}$. The electron-impurity scattering time $\tau$ and the Fermi velocity $v_F$ define the mean free path $\ell=v_F\tau$. The commutator in Eq.~(\ref{eqn:Eilenberger}) includes the convolution over time. Thus, the time derivative~$\partial_t$, the superconducting gap~$\Delta$, and the perturbation $\check{V}$ should be understood as integral operators in the time variables with kernels $\delta'(t-t')$, $\Delta(\vec{r})\,\delta(t-t')$, and $\check{V}(\vec{r},t,\vec{n})\,\delta(t-t')$, respectively.

In Keldysh space, $\partial_t$ and $\Delta$ are proportional to the $2\times2$ unit matrix. The same is true for a classical perturbation $V^\mathrm{c}$, in which case the left-lower corner of the Green's function $\hat{g}^Z=0$. However, the goal of this subsection is to evaluate the three components (retarded, advanced, and Keldysh) of the electronic polarization operator, $\Pi^{R,A,K}_{ijkl}(\vec{r},\vec{r}',\omega)$, needed in Eq.~(\ref{eqn:Q_ph_q}) for the cooling power.
%(here $\vec{r}$ denotes the center of mass coordinate, just like in the electronic quasiclassical Green's function~$\check{g}$). 
%
In a disordered superconductor, the calculation of the polarization operator involves summation of ladder diagram series, which is rather cumbersome~\cite{Sergeev1986}; in addition, here we are interested in a proximity system, lacking translational invariance. A more convenient way, equivalent in the quasiclassical limit $p_F\ell\gg1$, is to calculate the response of the electronic stress $\sigma_{ij}$ to an applied external classical strain $u_{ij}^\mathrm{c}$ using the Eilenberger equation~\cite{Narozhny1999}. Indeed, the average stress tensor (\ref{eqn:stress_tensor}) in terms of the quasiclassical Keldysh-Green's functions reads
\begin{equation}
	\sigma_{ij}(\vec{r},t)=\frac{\pi{N}_0}2\,\frac{p_F^2}m\left\langle
	\left(n_in_j-\delta_{ij}/3\right)
	\Tr\hat{g}^K(t,t;\vec{r},\vec{n})\right\rangle_\vec{n},
\end{equation}
where $N_0$ is the normal density of states at the Fermi level per spin projection. If the Green's function  is found to the first order in the perturbing stress, the result determines the retarded component of the polarization operator, $\Pi_{ijkl}^R$, since the latter coincides with the Kubo susceptibility (up to the sign). To find the advanced Keldysh components of the polarization operator, one has to include the quantum component $u_{ij}^\mathrm{q}$ of the strain. Thus, one has to consider the perturbation~$\check{V}$ with the following structure:
\begin{align}
\check{V}(t,t';\vec{r},\vec{n})={}&{}-\frac{p_F^2}m\left(n_in_j-{\delta_{ij}}/3\right)\delta(t-t')\,e^{-i\omega{t}}\nonumber\\
&{}\times\left(\begin{array}{cc} u_{ij}^\mathrm{c}(\vec{r}) & u_{ij}^\mathrm{q}(\vec{r}) \\  u_{ij}^\mathrm{q}(\vec{r}) & u_{ij}^\mathrm{c}(\vec{r}) \end{array}\right),
\end{align}
and calculate the response of $\check{g}$ to the first order in this perturbation from Eq.~(\ref{eqn:Eilenberger}). The three components of the electronic polarization operator can then be determined as~\cite{Narozhny1999}
\begin{widetext}\begin{subequations}\begin{align}
&\Pi^R_{ij,kl}(\vec{r},\vec{r}',\omega)=
-\frac{\pi{N}(0)}2\,\frac{p_F^2}m\left\langle\left(n_in_j-\frac{\delta_{ij}}3\right)
\Tr\frac{\delta\hat{g}^K(t,t;\vec{r},\vec{n})\,e^{i\omega{t}}}{\delta{u}_{kl}^\mathrm{cl}(\vec{r}')}\right\rangle_\vec{n},\\
&\Pi^A_{ij,kl}(\vec{r},\vec{r}',\omega)=
-\frac{\pi{N}(0)}2\,\frac{p_F^2}m\left\langle
\left(n_in_j-\frac{\delta_{ij}}3\right)
\Tr\frac{\delta(\hat{g}^R+\hat{g}^A)(t,t;\vec{r},\vec{n})\,e^{i\omega{t}}}{\delta{u}_{kl}^\mathrm{q}(\vec{r}')}\right\rangle_\vec{n},\\
&\Pi^K_{ij,kl}(\vec{r},\vec{r}',\omega)=
-\frac{\pi{N}(0)}2\,\frac{p_F^2}m\left\langle
\left(n_in_j-\frac{\delta_{ij}}3\right)
\Tr\frac{\delta(\hat{g}^K+\hat{g}^Z)(t,t;\vec{r},\vec{n})\,e^{i\omega{t}}}{\delta{u}_{kl}^\mathrm{q}(\vec{r}')}\right\rangle_\vec{n}.
\end{align}\end{subequations}\end{widetext}
Note that in the presence of the quantum component $u_{ij}^\mathrm{q}$, the lower left corner of Eq.~(\ref{eq:checkg=}), $\hat{g}^Z\neq0$.

We assume to be in the dirty limit, $\omega\tau\ll{1}$, $\Delta\tau\ll{1}$, so the unperturbed solution (at $\check{V}=0$) has the following angular structure~\cite{Usadel1970}:
\begin{equation}
\begin{split}
\check{g}(t,t';\vec{r},&\vec{n})=\int\frac{d\epsilon}{2\pi}\,e^{-i\epsilon(t-t')}
\big[\check{g}_0(\epsilon;\vec{r})-
\\
&-v_F\tau\vec{n}\cdot
\check{g}_0(\epsilon;\vec{r})\,\vec\nabla\check{g}_0(\epsilon;\vec{r})+O((\ell\nabla{g}_0)^2)\big].
\end{split}
\end{equation}
The isotropic part $\check{g}_0(\epsilon;\vec{r})$ satisfies the Usadel equation~\cite{Usadel1970}  which should be solved in each specific geometry of the proximitized system. In this subsection, we will assume that $\check{g}_0(\epsilon;\vec{r})$ is known.
Constraint (\ref{eqn:gg=1}) for $\check{g}$ implies
\begin{equation}\label{eq:g0g0=1}
\check{g}_0(\epsilon;\vec{r})\,\check{g}_0(\epsilon;\vec{r})
=\check{1}_{4\times4}+O((\ell\nabla{g}_0)^2).
\end{equation}
Usually, in the dirty limit it is sufficient to work with the Usadel equation without invoking the full Eilenberger equation~(\ref{eqn:Eilenberger}) at all. It is the angular structure of the perturbation, $n_in_j-\delta_{ij}/3$, proportional to the second spherical harmonics, that obliges us to work with Eq.~(\ref{eqn:Eilenberger}).

Let us first assume that the perturbation is a smooth function of space and time. The linear in $\check{V}$ correction, $\delta\check{g}$, to the leading order in $\tau\Delta$, $\tau\partial_t\check{V}$, $\ell\nabla\check{V}$, satisfies
%\begin{equation}
%[\check{g}_0,\delta\check{g}]+[\langle\delta\check{g}\rangle_\vec{n},\check{g}_0]
%=-2i\tau[\check{V},\check{g}_0],\quad
%\check{g}_0\,\delta\check{g}+\delta\check{g}\,\check{g}_0=0,
%\end{equation}
\begin{eqnarray}
&&[\check{g}_0,\,\delta\check{g}-\langle\delta\check{g}\rangle_\vec{n}]
=-2i\tau[\check{V},\check{g}_0],\label{eq:linEilendiff}\\
&&\check{g}_0\,\delta\check{g}+\delta\check{g}\,\check{g}_0=0,\label{eqn:lingg=1}
\end{eqnarray}
obtained by linearizing Eqs.~(\ref{eqn:Eilenberger}) and (\ref{eqn:gg=1}), respectively. Angular averaging gives $\langle\delta\check{g}\rangle_\vec{n}=0$ because $\langle\check{V}\rangle_\vec{n}=0$. Adding up (\ref{eq:linEilendiff}) and (\ref{eqn:lingg=1}), multiplying by $\check{g}_0$ on the left, and using Eq.~(\ref{eq:g0g0=1}), we obtain the response, local in space on the scale $\ell$:
\begin{equation}
\delta\check{g}=i\tau\left(\check{V}-\check{g}_0\check{V}\check{g}_0\right)
\left[1+O(\tau\Delta,\ell\nabla)\right].
\end{equation}
Let us now consider $u_{ij}(\vec{r})\propto e^{i\vec{q}\vec{r}-i\omega{t}}$ assuming $\omega\tau\ll{1}$, $\Delta\tau\ll{1}$, but not $q\ell\ll{1}$. As just seen, the spatial scale of the nonlocality in the response is~$\ell$, while $\check{g}_0$, found from the Usadel equation, depends on $\vec{r}$ on a longer scale. Then, to find the response at $q\sim{1}/\ell$, one can neglect the $\vec{r}$ dependence of $\check{g}_0$ and seek the correction $\delta\check{g}$ in the form $\delta\check{g}(\vec{r},\vec{n})=\delta\check{g}(\vec{n})\,e^{i\vec{q}\vec{r}}$.
%\begin{equation}
%\delta\check{g}(\vec{r},\vec{n})=
%\langle\delta\check{g}\rangle_\vec{n}\,e^{i\vec{q}\vec{r}}
%+\delta\check{g}_1(\vec{n})\,e^{i\vec{q}\vec{r}}.
%\end{equation}
The linearized Eilenberger equation becomes
\begin{equation}
\label{eqn:linEilenball}
[\check{g}_0,\,\delta\check{g}-\langle\delta\check{g}\rangle_\vec{n}]
+2i\ell(\vec{q}\vec{n})\,\delta\check{g}
=2i\tau[\check{g}_0,\check{V}],
\end{equation}
and the correction again satisfies Eq.~(\ref{eqn:lingg=1}).
Multiplying it with 1 and $(\vec{q}\vec{n})$, averaging over the angles, and using $\langle\check{V}\rangle_\vec{n}=0$, $\langle\vec{n}\check{V}\rangle_\vec{n}=0$, we obtain $\langle(\vec{q}\vec{n})\,\delta\check{g}\rangle_\vec{n}=0$, $\langle(\vec{q}\vec{n})^2\delta\check{g}\rangle_\vec{n}=0$.
Adding up Eq.~(\ref{eqn:lingg=1}) with its angular average subtracted, we obtain 
\begin{eqnarray}
&&\check{g}_0\,\delta\check{g}
+i\ell(\vec{q}\vec{n})\,\delta\check{g}
=\check{g}_0\langle\delta\check{g}\rangle_\vec{n}
+i\tau\check{g}_0\check{V}-i\tau\check{V}\check{g}_0,\\
&&i\ell(\vec{q}\vec{n})\check{g}_0\,\delta\check{g}+\delta\check{g}
=\langle\delta\check{g}\rangle_\vec{n}
+i\tau\check{V}-i\tau\check{g}_0\check{V}\check{g}_0,
\end{eqnarray}
where the second equation is obtained from the first by multiplying by $\check{g}_0$. This gives
\begin{align}
&\delta\check{g}=i\tau\,
\frac{1-i\ell(\vec{q}\vec{n})\check{g}_0}{1+\ell^2(\vec{q}\vec{n})^2}
\left(\frac{\langle\delta\check{g}\rangle_\vec{n}}{i\tau}+\check{V}
-\check{g}_0\check{V}\check{g}_0\right),\label{eq:deltag}\\
&\frac{\langle\delta\check{g}\rangle_\vec{n}}{i\tau}=
\left\langle\frac{\ell^2(\vec{q}\vec{n})^2}{1+\ell^2(\vec{q}\vec{n})^2}\right\rangle_\vec{n}^{-1}
\left\langle\frac{\check{V}-\check{g}_0\check{V}\check{g}_0}{1+\ell^2(\vec{q}\vec{n})^2}\right\rangle_\vec{n}.
\end{align}
Bearing in mind the structure of the strain-strain spectral function (\ref{eqn:spectral_D}), it is convenient to define three components of the polarization operator:
%\begin{subequations}\begin{align}
%&\Pi_l(\vec{q},\omega)=\frac{q_iq_jq_mq_n}{q^4}\,\Pi_{ijmn}(\vec{q},\omega),\quad
%\\
%&\Pi_{t\kappa}(\vec{q},\omega)=
%\frac{(e_{t\kappa}^iq_j+e_{t\kappa}^jq_i)(e_{t\kappa}^mq_n+e_{t\kappa}^nq_m)}{4q^2}\,\Pi_{ijmn}(\vec{q},\omega)
%\end{align}\end{subequations}
%with $\kappa=1,2$,
\begin{equation}
\Pi_\lambda(\vec{q},\omega)=\mathcal{T}^{\vec{q}\lambda}_{ij}\,\mathcal{T}^{\vec{q}\lambda}_{kl}\,\Pi_{ijkl}(\vec{q},\omega),\quad
\lambda=l,t1,t2,
\end{equation}
which can be found from Eq.~(\ref{eq:deltag}) for the following perturbations:
\begin{align}
&\check{V}_\lambda(t,t';\vec{r},\vec{n})=\delta(t-t')\,e^{i\vec{q}\vec{r}-i\omega{t}}\,\Phi_\lambda(\vec{n})\left(\begin{array}{cc} u_\lambda^\mathrm{c} & u_\lambda^\mathrm{q} \\  u_\lambda^\mathrm{q} & u_\lambda^\mathrm{c}
\end{array}\right),\\
\label{eqn:Phi_lambda}
&\Phi_l(\vec{n})=\frac{(\vec{q}\vec{n})^2}{q^2}-\frac{1}3,
\quad\Phi_{t1,t2}(\vec{n})=\frac{(\vec{q}\vec{n})(\vec{e}_{t1,t2}\vec{n})}{q}.
\end{align}
As a result,
\begin{widetext}
\begin{subequations}\begin{eqnarray}
\label{eqn:Pi^R/A}
\Pi^{R/A}_\lambda(\vec{q},\omega)=&&~2N_0\left(\frac{p_F^2}m\right)^2
\left[\mathcal{Y}_\lambda(0)+\frac{i\tau}8\,\mathcal{Y}_\lambda(q\ell)\int\limits_{-\infty}^\infty{d}\epsilon\,
\Tr\left\{\hat{g}^{R/A}_0(\epsilon_+)\,\hat{g}^K_0(\epsilon_-)
+\hat{g}^K_0(\epsilon_+)\,\hat{g}^{A/R}_0(\epsilon_-)\right\}\right],
\\
\label{eqn:Pi^K}
\Pi^{K}_\lambda(\vec{q},\omega)=&&~2N_0\left(\frac{p_F^2}m\right)^2
\left[\frac{i\tau}8\,\mathcal{Y}_\lambda(q\ell)\int\limits_{-\infty}^\infty{d}\epsilon\,
\Tr\left\{\hat{g}^{K}_0(\epsilon_+)\,\hat{g}^K_0(\epsilon_-)
-\hat{g}^{R-A}_0(\epsilon_+)\,\hat{g}^{R-A}_0(\epsilon_-)\right\}\right],
\end{eqnarray}\label{eqn:Pi=}\end{subequations}
\end{widetext}
where $\lambda=l,t1,t2$, $\epsilon_{\pm} = \epsilon\pm \omega/2$ and $\hat{g}_0^{R-A} = \hat{g}_0^{R}-\hat{g}_0^{A}$. The term with $\mathcal{Y}_\lambda(0)$ is not captured by the quasiclassical theory and is inserted by noting that the response at $\omega=0$ is determined by the Fermi sea, and thus is (i)~insensitive both to disorder and to superconductivity, and (ii)~is local on the spatial scale of the Fermi wavelength, so it can be evaluated for a clean Fermi gas at $q=0$~\cite{Narozhny1999}. The factors $\mathcal{Y}_\lambda(q\ell)$ coming from angular averages (see Appendix~\ref{appendix:qL_factors}) are given by 
\begin{subequations}\label{eqn:q_l/t}
\begin{align}
&\mathcal{Y}_l(\xi)=-\frac{\xi-(1+\xi^2/3)\arctan{\xi}}{3\xi^2(\xi-\arctan{\xi})},
\\
&\mathcal{Y}_{t1,t2}(\xi)=\frac{\xi(1+2\xi^2/3)-(1+\xi^2)\arctan{\xi}}{2\xi^5}.
\end{align}\end{subequations}
Remarkably, Eqs.~(\ref{eqn:Pi=}) have a separable form: the dependence on $\vec{q}$ and $\lambda$ is factorized from the rest which contains the frequency and coordinate dependence and all information about the superconductivity and the proximity effect. The $\vec{q},\lambda$ dependence is entirely contained in the factors $\mathcal{Y}_\lambda(q\ell)$, and is the same as calculated for a normal metal~\cite{Schmid1973}.

\subsection{Final expression for the cooling power}

Having obtained the expressions (\ref{eqn:Pi=}) for the polarization operator, we can  rewrite Eq.~(\ref{eqn:Q_ph_q}) for the cooling power per unit volume as follows:
\begin{equation}\begin{split}
Q = {}&{} \sum_\lambda\int\frac{d\vec{q}}{(2\pi)^3}
\int_{-\infty}^{\infty} \frac{d\omega}{2\pi}\, \frac\omega{4}\,
D^{R-A}_\lambda(\vec{q},\omega)\\
& {}\times
2N_0\left(\frac{p_F^2}m\right)^2
\left(\frac{i\tau}{8}\right)\mathcal{Y}_\lambda(q\ell)\, \mathcal{F}(\vec{R},\omega).
%\left\{\left[\Pi^R_\lambda(\vec{R},\vec{q},\omega)-\Pi^A_\lambda(\vec{R},\vec{q},\omega)\right]
%\coth\frac{\omega}{2T_\mathrm{ph}}-\Pi^K_\lambda(\vec{R},\vec{q},\omega)\right\}.
\end{split}
\end{equation}
Here $\mathcal{F}(\vec{R},\omega)$ denotes the factor
\begin{align}
\mathcal{F}={}&{}\int_{-\infty}^\infty{d}\epsilon\times{}\nonumber\\
&{}\times\Tr\bigg\{\left[\hat{g}^{R-A}_0(\epsilon_+)\coth\frac{\omega}{2T_\mathrm{ph}}-\hat{g}^K_0(\epsilon_+)\right]\,\hat{g}^K_0(\epsilon_-)\nonumber
\\
&\quad{}-\left[\hat{g}^K_0(\epsilon_+)\coth\frac{\omega}{2T_\mathrm{ph}} - \hat{g}^{R-A}_0(\epsilon_+)\right]\hat{g}^{R-A}_0(\epsilon_-)\bigg\},\label{eqn:F=}
\end{align}
whose frequency dependence comes from $\epsilon_\pm=\epsilon\pm\omega/2$ and the coordinate dependence from that of the Green's function $\check{g}_0(\vec{R},\epsilon_\pm)$ whose spatial argument is omitted in Eq.~(\ref{eqn:F=}) for brevity.
Since the $\vec{q}$ dependence of the electronic spectral function is solely contained in $\mathcal{Y}_\lambda(q\ell)$, the integration over $\vec{q}$ is straightforwardly performed by resolving the $\delta$~functions in Eq.~(\ref{eqn:Dlambda}), which yields
\begin{equation}
\label{eqn:heat_current}
Q = \frac{N_0\tau}{8\pi\rho_0}\left(\frac{p_F^2}m\right)^2\int\limits_{0}^{\infty}\frac{d\omega}{2\pi}
\sum_\lambda\frac{\omega^4}{c_\lambda^5}
%\mathcal{Y}_\lambda\!\left(\frac{\omega\ell}{c_\lambda}\right)
\mathcal{Y}_\lambda(\omega\ell/c_\lambda)\,
\mathcal{F}(\vec{R},\omega).
\end{equation}
This very general formula is the main result of this paper and it is applicable to variety of electronic systems including superconducting proximity structures. If the electron-electron relaxation is sufficiently fast we can assume the electrons to be in thermal equilibrium at temperature~$T_\mathrm{e}$ \cite{Pothier1997},
%\begin{equation}
%\hat{g}^K_0(\epsilon) =\left[\hat{g}^R_0(\epsilon) - \hat{g}^A_0(\epsilon)\right]\tanh\frac{\epsilon}{2T_\mathrm{e}},
%\end{equation}
so that $\hat{g}^K_0(\epsilon)=\hat{g}^{R-A}_0(\epsilon)\tanh[\epsilon/(2T_\mathrm{e})]$, then Eq.~(\ref{eqn:heat_current}) further simplifies adopting the form
%\begin{align}
%&Q_{\rm ph}(T_\mathrm{ph}, T_\mathrm{e}) = 
%\frac{N_0\tau}{8\pi\rho_0}\left(\frac{p_F^2}m\right)^2
%\int d\vec{R}\int\limits_{0}^{\infty}\frac{d\omega}{2\pi}\sum_\lambda\frac{\omega^4}{c_\lambda^5}\,\mathcal{Y}_\lambda\!\left(\frac{\omega\ell}{c_\lambda}\right)
%\nonumber\\ &\qquad\qquad\qquad\quad{}\times
%I(\vec{R},\omega)
%\left( \coth\frac{\omega}{2T_\mathrm{ph}} - \coth\frac{\omega}{2T_\mathrm{e}}\right) ,
%\end{align}
\begin{align}
Q(T_\mathrm{e}, T_\mathrm{ph}) = {}&{}
\frac{N_0\tau}{8\pi\rho_0}\left(\frac{p_F^2}m\right)^2
\int_{0}^{\infty}\frac{d\omega}{2\pi}\sum_\lambda\frac{\omega^4}{c_\lambda^5}\,\mathcal{Y}_\lambda(\omega\ell/c_\lambda)\nonumber\\
%%%
{}&{}\times
\left(\coth\frac{\omega}{2T_\mathrm{e}}-\coth\frac{\omega}{2T_\mathrm{ph}}\right)
\mathcal{I}(\vec{R},\omega),
\label{eqn:heat_current_eql}
\end{align}
where 
\begin{align}
	\label{eqn:I_general}
	\mathcal{I}(\vec{R},\omega)={}&{}2\int_{-\infty}^\infty{d}\epsilon\,
	\left[n_F(\epsilon_-)-n_F(\epsilon_+)\right]\times{}\nonumber
	\\
	&{}\times
	\Tr\left\{\hat{g}^{R-A}_0(\vec{R},\epsilon_+)\,\hat{g}^{R-A}_0(\vec{R},\epsilon_-)\right\},
\end{align}
and $n_F(\epsilon)=[\exp(\epsilon/T_\mathrm{e})+1]^{-1}$ is the Fermi distribution. The  whole information about the electronic properties of the system is contained in the function $\mathcal{I}(\vec{R},\omega)$. Essentially, our task from now on is to calculate it for various systems. Another important quantity we are interested in is the thermal conductance per unit volume:
%\begin{widetext}
\begin{equation}
\label{eqn:G_definition}
K(T)=\frac{\partial Q(T_\mathrm{e},T)}{\partial T_\mathrm{e}}\bigg|_{T_\mathrm{e}=T}.
\end{equation}
%\end{widetext}
%\textcolor{red}{(Denoted by $K$ to avoid confusion with the electrical conductance which will appear later.)}
Eqs.~(\ref{eqn:heat_current_eql}) and (\ref{eqn:G_definition}) yield an expression for the thermal conductance of the same form as Eq.~(\ref{eqn:heat_current_eql}), but with a replacement
\begin{equation}
\coth\frac{\omega}{2T_\mathrm{e}} - \coth\frac{\omega}{2T_\mathrm{ph}}\to
\frac{\omega}{2T^2\sinh[\omega/(2T)]}.
\end{equation}
%\begin{widetext}
%\begin{equation}
%\label{eqn:thermal_conductance}
%\begin{split}
%G_{\rm ph}(T)\quad{} &= -\frac{N_0\tau}{8\pi\rho_0}\left(\frac{p_F^2}m\right)^2\frac{1}{T^2}\int d\vec{R}\int\limits_{-\infty}^{\infty}\frac{d\omega}{2\pi}\bigg\{\sum_\lambda\frac{\omega^5}{c_\lambda^5}\mathcal{Y}_\lambda(\omega\ell/c_\lambda)
%%%
%\frac{ e^{\omega/T}}{(e^{\omega/T}-1)^2}\times
%\\
%&{}\times 2\int\limits_{-\infty}^\infty{d}\epsilon
%\Tr\bigg\{\left[\hat{g}^R_0(\vec{R};\epsilon_+)-\hat{g}^A_0(\vec{R};\epsilon_+)\right]\big[\hat{g}^R_0(\vec{R};\epsilon_-)-\hat{g}^A_0(\vec{R};\epsilon_-)\big]
%\left[n_F(\epsilon_+)-n_F(\epsilon_-)\right]\bigg\},
%\end{split}
%\end{equation}
%\end{widetext}
%This formula has the same level of generality as Eq.~(\ref{eqn:heat_current_eql}) and according to its definition the thermal conductance depends on only one temperature,~$T$.
In the following section we shall make use of the developed formalism to calculate the electron-phonon cooling power in various electronic systems.

\section{Application to specific structures}
\label{sec:results}
\subsection{Bulk normal metal}
\label{sec:results_N}

%%%%%%%%%%%%%%%%%%%%%%%%%%%
\begin{figure}[b]
	\centering
	\includegraphics [width=\columnwidth] {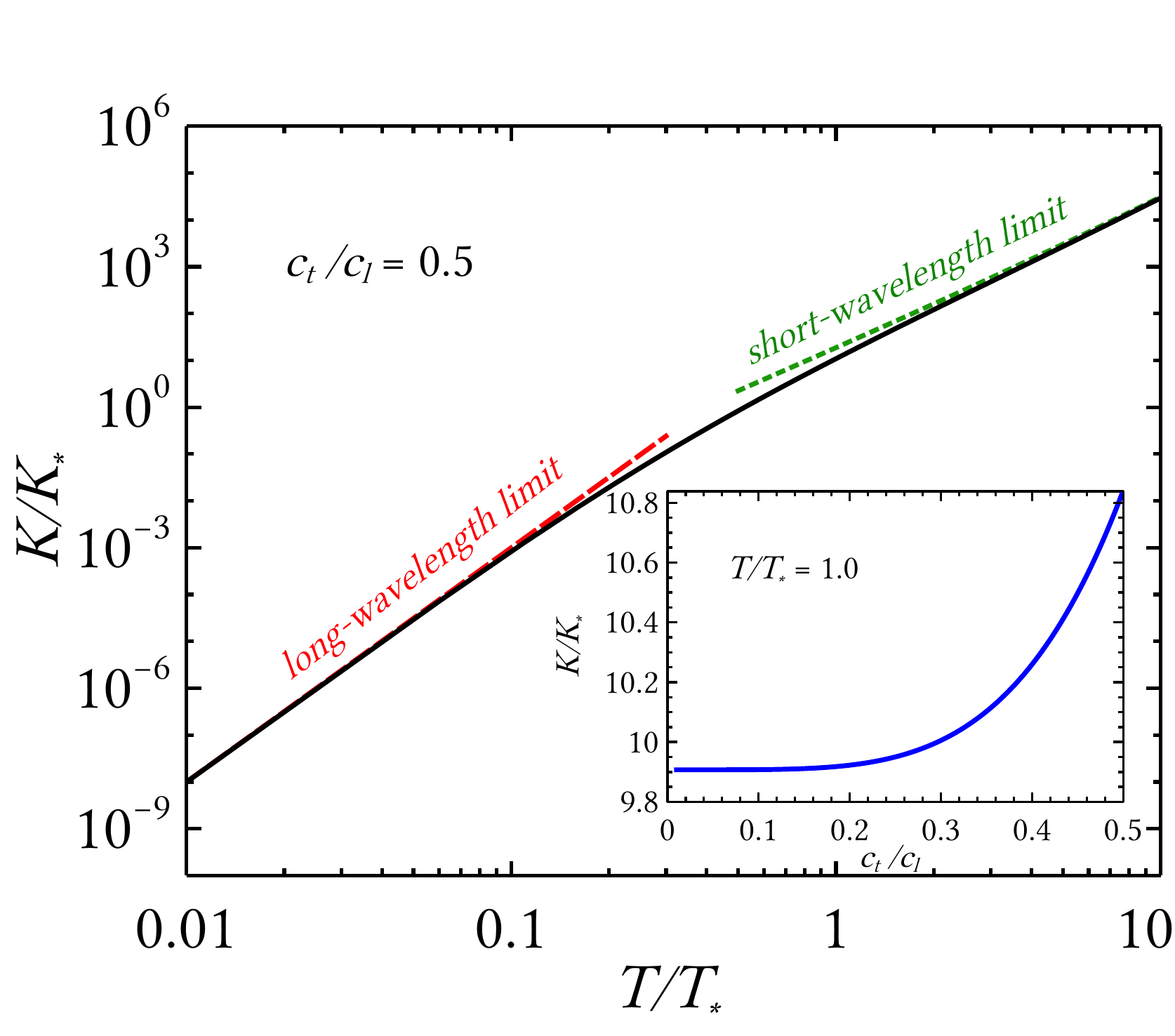}
	\caption{(Color online) Electron-phonon thermal conductance per unit volume in a bulk normal metal as a function of temperature $T$ for $c_l/c_t=0.5$. The dotted and dashed lines correspond to the high- and low-$T$ limits, respectively. The inset shows the same quantity as a function of $c_{t}/c_{l}$ for the temperature $T/T_* = 1.0$.} 
	\label{fig:GN}
\end{figure}
%%%%%%%%%%%%%%%%%%%%%%%%%%%

In a bulk normal metal, the retarded and advanced component of $\check{g}$ are just $\hat{g}^{R/A}=\pm\hat{\tau}_3$,  so the function $\mathcal{I}(\omega)$ from Eq.~(\ref{eqn:I_general}) simply reads $\mathcal{I}(\omega)=16\,\omega$.
Plugging it into Eq.~(\ref{eqn:heat_current_eql}), we obtain the cooling power per unit volume in the form $Q(T_\mathrm{e})-Q(T_\mathrm{ph})$ with $Q(T)$ given by:
\begin{equation}\label{eq:Qnormal}
Q(T)=\frac{2N_0\tau}{\pi^2\rho_0}\left(\frac{p_F^2}m\right)^2\sum_\lambda\int\limits_0^\infty d\omega\,\frac{\omega^5}{c_\lambda^5}\,
\frac{\mathcal{Y}_\lambda(\omega\ell/c_\lambda)}{e^{\omega/T}-1}.
\end{equation}
Since the functions $\mathcal{Y}_{\lambda}(\xi)$ [see Eq.~(\ref{eqn:q_l/t})] are rather complicated, the integral should be calculated numerically. Still, it simplifies in two limiting cases.

At low temperatures, $T\ll{c_\lambda}/\ell$, we employ Eqs.~(\ref{eqn:Ql0}) and (\ref{eqn:Qt0}) for $\mathcal{Y}_\lambda(\xi)$, and arrive at the well-known $T^6$ dependence~\cite{Schmid1973, Sergeev1986},
\begin{equation}
\label{eqn:GN_long}
Q(T) = \frac{32\pi^4}{945}\frac{N_0c_t\tau}{\rho_0\ell^6}\left(\frac{p_F^2}m\right)^2\left(1+\frac{2}{3}\,\frac{c_t^5}{c_l^5}\right)\frac{T^6}{T_*^6},
\end{equation}
where the crossover temperature $T_*\equiv{c}_t/\ell$. 
%\textcolor{red}{(Asterisk in the subscript in order to write powers.)}
At high temperatures, $T\gg{c}_\lambda/\ell$, we use Eqs.~(\ref{eqn:Qlinf}) and (\ref{eqn:Qtinf}) for $\mathcal{Y}_\lambda(\xi)$ we end up with the following expression for the cooling power:
\begin{equation}
Q(T) = \frac{N_0c_t\tau}{\rho_0\ell^6}\left(\frac{p_F^2}m\right)^2
\left(\frac{3\,\zeta(5)}{3\pi}\,\frac{c_t^4}{c_l^4}\,\frac{T^5}{T_*^5}
+\frac{4\pi^2}{45}\frac{T^4}{T_*^4}\right),\label{eqn:QN_short}
\end{equation}
where $\zeta(x)$ is the Riemann zeta function. In the clean limit, $\tau,\ell\rightarrow \infty$ with $\ell/\tau=v_F$, the second term $\propto{T}^4$ vanishes, while the first one gives the standard $Q(T)\propto{T}^5$ result for clean metals \cite{Price1982,Wellstood1994}.

The thermal conductance per unit volume, $K(T)=dQ(T)/dT$, is shown in Fig.~\ref{fig:GN}.  As can be seen from Eq.~(\ref{eq:Qnormal}), $K(T)$ has a natural unit
\begin{equation}
	K_*=\frac{2N_0\tau}{\pi^2\rho_0\ell^5}\left(\frac{p_F^2}m\right)^2,
\end{equation}
so that $K(T)/K_*$ is a dimensionless function of two dimensionless parameters $T/T_*\equiv{T}\ell/c_t$ and $c_t/c_l$ (note that $c_t/c_l<1/\sqrt{2}$~\cite{Landafshitz7}). We numerically evaluate the integral in Eq.~(\ref{eq:Qnormal}) and plot $K/K_*$ as a function of $T/T_*$ for $c_t/c_l = 0.5$ in Fig.~\ref{fig:GN}.
%\textcolor{red}{(Replot the figure for $c_t/c_l=0.1, 0.3, 0.5$.)}
The low- and high-$T$ limits are indicated by the dashed and dotted lines, respectively. Their validity depends on $c_t/c_l$, but roughly speaking the two limits are reached for $T/T_*<0.1$ and $T/T_*>1$. The inset shows the cooling power as a function of the $c_t/c_l$ ratio at temperature $T/T_* = 1.0$. Estimation of the crossover temperature in copper with $\ell=10$~nm is $T_*\approx 1.8$~K. The values taken for the longitudinal and transverse speed of sound are, respectively, $c_l = 4.8$~km/s and $c_t =$~2.3 km/s.

%\textcolor{red}{(List other parameters for copper that you used to get this estimate. $c_l=4.8\:\mbox{km/s}$, $c_t=2.3\:\mbox{km/s}$?)}

%\clearpage
\subsection{Bulk BCS superconductor}
In a superconductor, the retarded and advanced components of the quasiclassical Keldysh Green's function $\hat{g}^{R/A}(\epsilon)$ can be parameterized in terms of the normal, $g$, and the anomalous, $f,f^\dagger$, Green's functions (we omit the superscripts $R,A$ for compactness):
\begin{equation}
\label{eqn:GF_superconductor}
\hat{g}(\epsilon)=
\left(\begin{array}{cc}
g(\epsilon) &f(\epsilon)\\ 
f^{\dagger}(\epsilon)&-g(\epsilon)
\end{array}\right),\quad g^2(\epsilon)+f(\epsilon)\,f^{\dagger}(\epsilon) =1.
\end{equation}
Assuming the superconducting gap to be real, $\Delta=\Delta^*$, we also have $f=f^{\dagger}$. Substituting the parametrization~(\ref{eqn:GF_superconductor}) into Eq.~(\ref{eqn:I_general}) and employing the relation $\hat{g}^{R} = -\hat{\tau}_3(\hat{g}^A)^{\dagger}\hat{\tau}_3$~\cite{Rammer1986}, we obtain
\begin{align}\label{eqn:I_S}
\mathcal{I}(\omega)={}&{}16\int\limits_{-\infty}^{\infty}d\epsilon\left[n_F(\epsilon_-,T)-n_F(\epsilon_+,T)\right]\nonumber\\
&{}\times[\Re{g}^R(\epsilon_+)\Re{g}^R(\epsilon_-)-\Im{f}^R(\epsilon_+)\Im{f}^R(\epsilon_-)].
\end{align}
We note that this expression is rather general and will be applied to the bulk homogeneous superconductor immediately below, as well as to other proximity structures in the following subsections. We also note that since the Green's functions depend on the electronic temperature $T_\mathrm{e}$ via the superconducting gap, the cooling power can no longer be represented in the form $Q(T_\mathrm{e})-Q(T_\mathrm{ph})$. In the following, we will focus on the thermal conductance per unit volume, Eq.~(\ref{eqn:G_definition}), which depends only on one temperature. It is given by
\begin{align}
\frac{K(T)}{K_*}={}&{}\int_0^\infty\frac{\omega^5\,\mathcal{I}(\omega)\,d\omega}{64T^2\sinh^2[\omega/(2T)]}\nonumber\\
&{}\times\left[\frac{\mathcal{Y}_l(\omega\ell/c_l)}{(c_l/\ell)^5}
+2\,\frac{\mathcal{Y}_t(\omega\ell/c_t)}{(c_t/\ell)^5}\right].\label{eqn:G_S}
\end{align}

In a bulk homogeneous superconductor, the quasiclassical Green's functions are given by
\begin{equation}
\label{eqn:g_S}
g(\epsilon)=\frac{-i\epsilon}{\sqrt{\Delta^2-\epsilon^2}},\quad
f(\epsilon)= f^{\dagger}(\epsilon)= 
\frac{\Delta}{\sqrt{\Delta^2-\epsilon^2}}.
\end{equation}
The retarded/advanced Green's function is obtained by the substitution $\epsilon\rightarrow\epsilon\pm i\eta$. The broadening parameter $\eta$ can be taken to be infinitesimal, or finite, describing level broadening due to some relaxation processes~ \cite{Dynes1978,Dynes1984}. The temperature dependence of the superconducting gap is assumed to be \cite{Gap}
\begin{equation}
\Delta(T)=\Delta_0\tanh(1.74\sqrt{T_c/T-1}),
\end{equation}
where $\Delta_0$ is the superconducting gap at zero temperature and $T_c$ is the critical temperature.

%%%%%%%%%%%%%%%%%%%%%%
\begin{figure}[t]
	\centering
	\includegraphics[width=8cm]{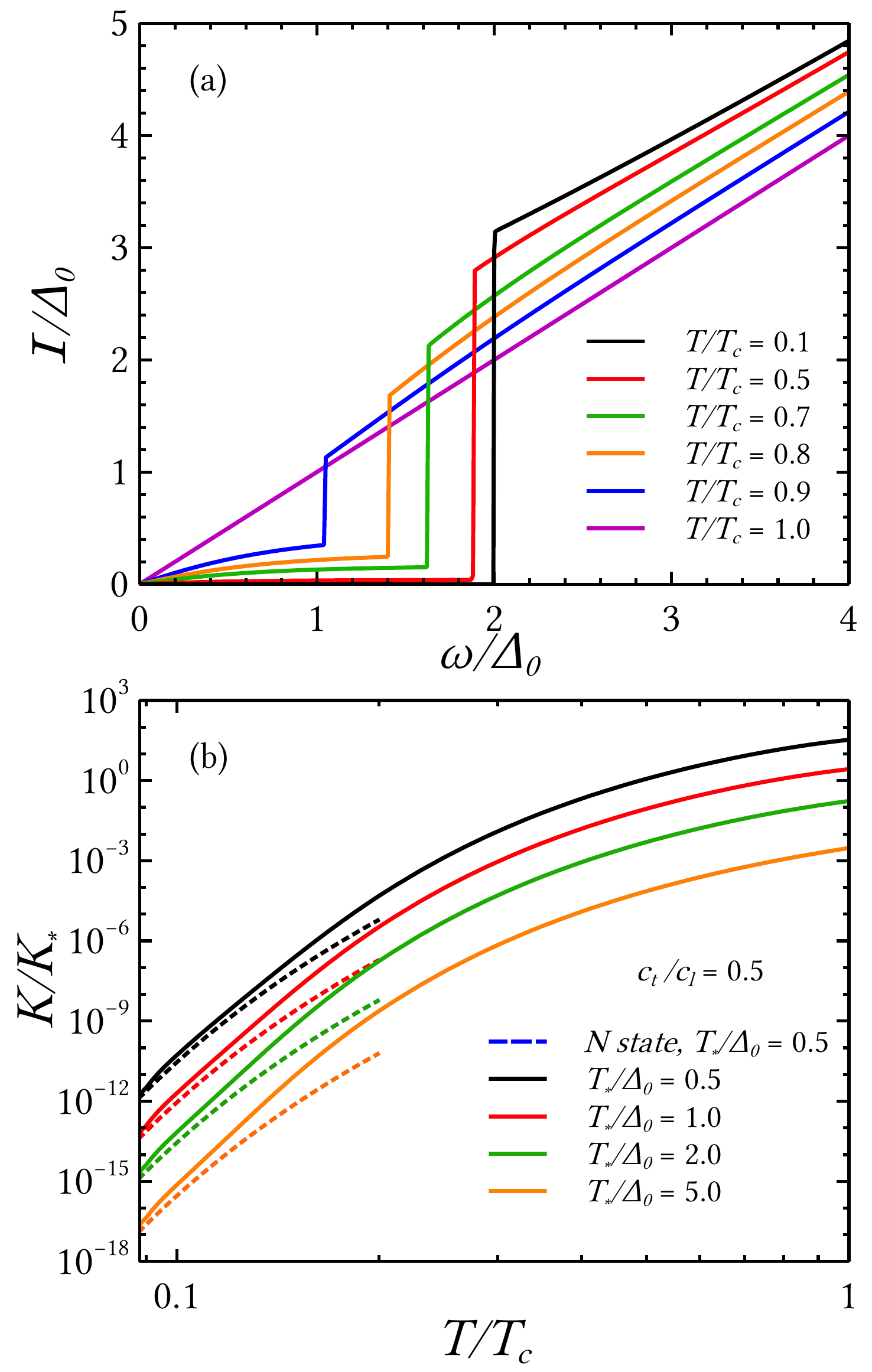}
	\caption{(Color online) (a) The $\mathcal{I}(\omega)$ function of a bulk BCS superconductor for various temperatures. At $T=T_c$ the normal state result $\mathcal{I}(\omega)=16\,\omega$ is recovered.
		 (b) Electron-phonon thermal conductance per unit volume of a bulk superconductor as a function of temperature $T$ for $c_t/c_l=0.5$ and several values of $T_*/\Delta_0$. The dashed blue line corresponds to the normal case with $T_*/\Delta_0=0.5$ (whose only role in the normal state is to set the scale of the horizontal axis). The dotted lines show the low-temperature asymptotics~(\ref{eqn:KlowT}).
 	}
	\label{fig:IG_BCS}
\end{figure}
%%%%%%%%%%%%%%%%%%%%%

Plugging the Green's functions~(\ref{eqn:g_S}) into Eq.~(\ref{eqn:I_S}), we obtain $\mathcal{I}(\omega)$ of a bulk BCS superconductor, 
\begin{align}
\mathcal{I}(\omega)=16\int_{-\infty}^\infty{d}\epsilon
\left[n_F(\epsilon_-,T)-n_F(\epsilon_+,T)\right]\nonumber\\
{}\times
\frac{\theta(|\epsilon_+|-\Delta)\,\theta(|\epsilon_-|-\Delta)}{\sqrt{\epsilon_+^2-\Delta^2}\sqrt{\epsilon_-^2-\Delta^2}}\nonumber\\
{}\times\left(|\epsilon_+\epsilon_-|-\Delta^2\sign\epsilon_+\epsilon_-\right)
\label{eqn:IbulkBCS=}
\end{align}
[$\theta(x)$ is the Heaviside step function],
illustrated in Fig.~\ref{fig:IG_BCS}(a) for different temperatures. At  $T=T_c$, the normal state dependence $\mathcal{I}(\omega)=16\,\omega$ is recovered. The main difference between the normal and superconducting cases is the presence of a gap in $\mathcal{I}(\omega)$ at $\omega<2\Delta$.
%Apparently, this is a consequence of the presence of the superconducting gap in the quasiparticle spectrum.
With increasing temperatures this gap shrinks, and it is not empty anymore due to thermal quasiparticle population (see the blue line in Fig.~\ref{fig:IG_BCS}(a) that corresponds to $T/T_c=0.8$).
At $T\ll\Delta$,
\begin{subequations}\begin{align}
&\mathcal{I}(0<\omega<2\Delta)=16\sqrt{\frac{2\pi\omega\Delta{T}}{\omega+2\Delta}}\,
\left(1-e^{-\omega/T}\right)e^{-\Delta/T},\label{eqn:IbulkBCSlowT}\\
&\mathcal{I}(\omega=2\Delta+0^+)=16\pi\Delta.
\end{align}\end{subequations}

The thermal conductance per unit volume $K(T)$, besides $T/T_*$ and $c_l/c_t$, now depends on another dimensionless parameter $T_*/\Delta_0$. For aluminum with the electronic mean free path of $\ell = 10$~nm, $T_*/\Delta_0\approx{1.1}$. The values for the longitudinal and transverse speed of sound are taken $c_l = 6.4$~km/s and  $c_t = 3.0$~km/s, respectively.
%\textcolor{red}{(List other parameters for aluminum that you used to get this estimate. $c_l=6.4\:\mbox{km/s}$, $c_t=3.0\:\mbox{km/s}$?)}
Plugging Eq.~(\ref{eqn:IbulkBCS=}) into Eq.~(\ref{eqn:G_S}) and evaluating the integral numerically, we show $K(T)$ in Fig.~\ref{fig:IG_BCS}(b) for $c_t/c_l=0.5$ and several values of $T_*/\Delta_0$. Increase of $T_*/\Delta_0$ almost does not change the shape of the curves just shifting them along the vertical axis. 
At low temperatures, the cooling power in a BCS superconductor is exponentially suppressed in comparison to the normal state [see the dashed blue line in Fig.~\ref{fig:IG_BCS}(b)] due to the presence of the superconducting gap:
\begin{equation}\label{eqn:KlowT}
\frac{K(T\ll\Delta)}{K_*}=\frac{693\pi}{32}\,\zeta\!\left(\frac{13}2\right)
\left(1+\frac{2}{3}\,\frac{c_t^5}{c_l^5}\right)\frac{T^5}{T_*^5}\,e^{-\Delta/T}.
\end{equation}
This difference diminishes at higher temperatures, and finally at $T=T_c$ the normal case is recovered.

\subsection{Thin SIN contact}

%%%%%%%%%%%%%%%%%%%%%%%%%
\begin{figure}[b]
	\centering
	\includegraphics[width=\columnwidth]{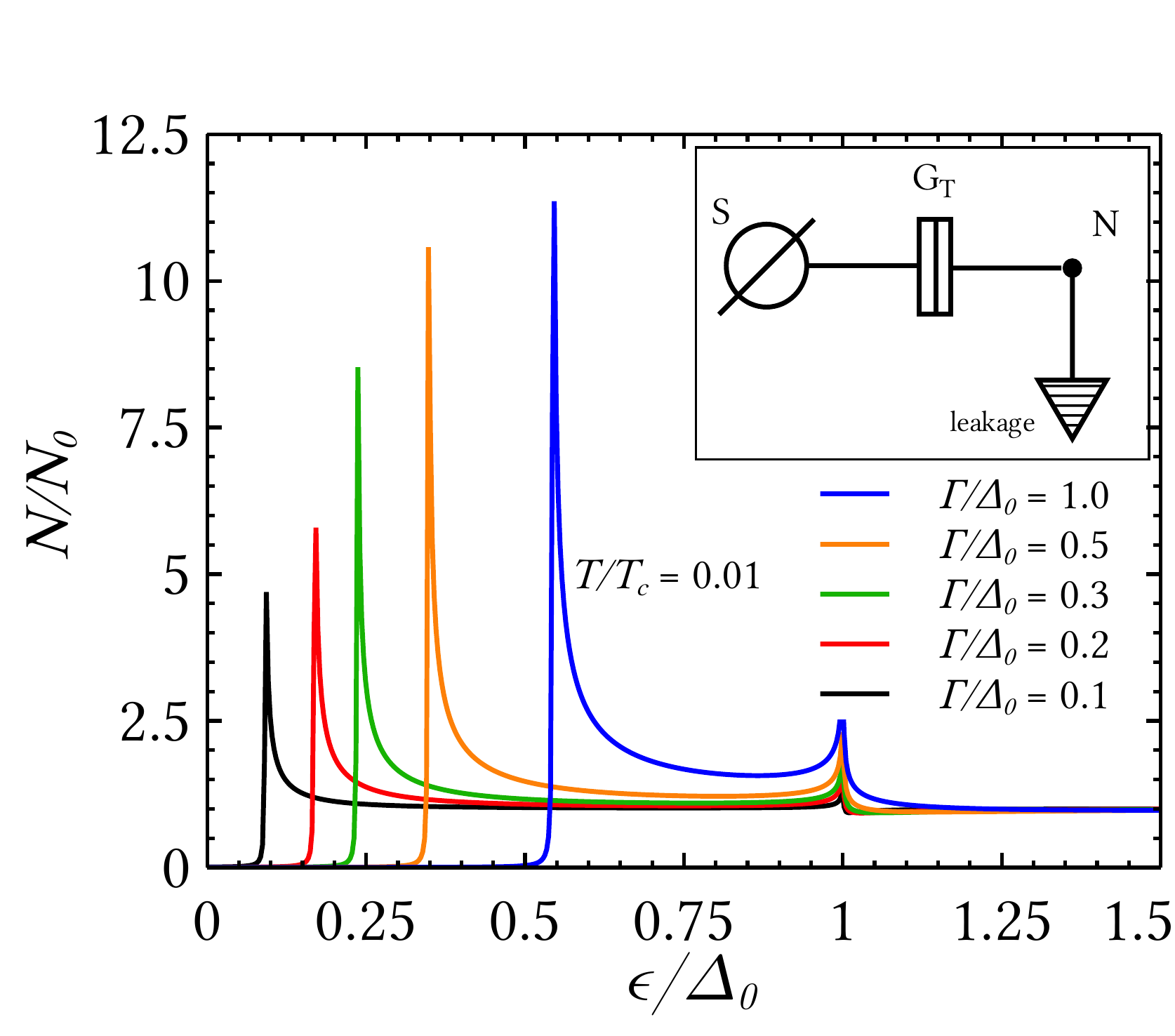}
	\caption{(Color online) The density of state $N(\epsilon)$ in a thin normal metal coupled to a massive superconducting lead via a tunnel contact for several values of $\Gamma/\Delta_0$, and temperature $T/T_c=0.01$. The Dynes broadening parameter is $\eta/\Delta_0=10^{-3}$. The inset shows a schematic view of the structure.}
	\label{fig:SN_DOS}
\end{figure}
%%%%%%%%%%%%%%%%%%%%%%%%%

Let us now consider a simple $\S\N$ proximity structure consisting of a small island of normal metal coupled to a massive superconducting electrode via a weak tunnel contact. Then we can neglect suppression of superconductivity by the inverse proximity effect in the superconductor, and focus on the proximity effect in the normal part. In the zero-dimensional limit it can be described by the quantum circuit theory~\cite{Nazarov1999}. The quasiclassical Green's function of the normal metal $\hat{g}_N^{R/A}$ satisfies the zero-dimensional analog of the Usadel equation~\cite{Nazarov1999, Boerlin2002, Morten2006, noteUsadel0D}:
\begin{equation}
\label{eqn:Usadel0D}
-i\epsilon\left[\hat{\tau}_3, \hat{g}_N^{R/A}(\epsilon)\right] + \Gamma
\left[\hat{g}_S^{R/A}(\epsilon),\hat{g}_N^{R/A}(\epsilon)\right] = 0,
\end{equation}
where $\hat{g}_S^{R/A}(\epsilon)$ is the solution for a homogeneous BCS superconductor given in Eq.~(\ref{eqn:g_S}) and the first commutator denotes the so-called leakage of coherence [see the inset in Fig.~\ref{fig:SN_DOS}]. $\Gamma$~is half of the rate of electron escape from the island into the bulk electrode, related to the tunnel contact conductance $G$ via $G=4e^2N_0\mathcal{V}\Gamma$, where $\mathcal{V}$ is the island volume, so that $1/(N_0\mathcal{V})$ is the electronic orbital mean level spacing in the island. $\Gamma$~must be small compared to the island Thouless energy $E_\mathrm{Th}$, defined as the inverse time needed for an electron to cross the island, in order for the island to be in the zero-dimensional limit; at the same time, we need $\Gamma\gg1/(N_0\mathcal{V})$ for the Coulomb blockade effects to be negligible~\cite{Aleiner2002}.
Eq.~(\ref{eqn:Usadel0D}) can also describe a planar structure when both $\mathcal{V}$ and $G$ are proportional to the contact area, while $\Gamma$ is independent of the area; in this geometry the normal layer thickness~$d$ should be larger than the mean free path but small enough so that the time needed for an electron to travel the distance~$d$ is smaller than $1/\Gamma$ (see the next subsection for more details). 

The solution of Eq.~(\ref{eqn:Usadel0D}) reads
\begin{align}
\label{eqn:g_SN}
&\hat{g}_N(\epsilon)= \frac{a\hat{\tau}_3+b\hat{\tau}_1}{\sqrt{a^2+b^2}}, \\
&a\equiv-i\epsilon\left(1+\frac{\Gamma}{\sqrt{\Delta^2-\epsilon^2}}\right),\quad
b\equiv\frac{\Gamma\Delta}{\sqrt{\Delta^2-\epsilon^2}},\nonumber
\end{align}
and the retarded and advanced functions are obtained by substituting $\epsilon\to\epsilon\pm{i}\eta$, as in the previous subsection.
%

%%%%%%%%%%%%%%%%%%%%%%%%%
\begin{figure}[b]
	\centering
	\includegraphics[width=8cm]{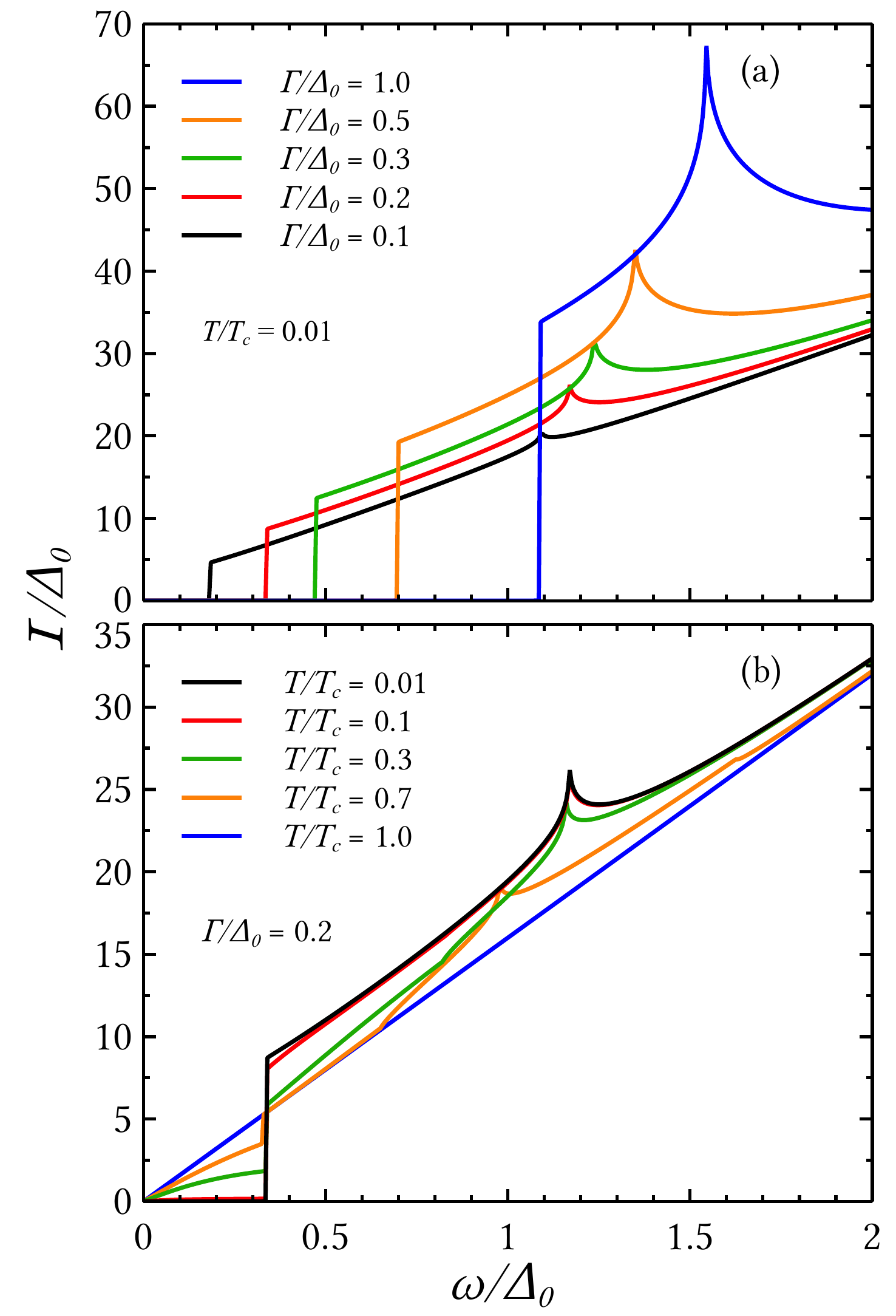}
	\caption{(Color online) $\mathcal{I}(\omega)$ function in a thin normal metal coupled to a massive superconducting lead for (a)~different values of $\Gamma/\Delta_0$, and temperature $T/T_c=0.01$, (b) different temperatures and $\Gamma/\Delta_0=0.2$. The broadening parameter $\eta/\Delta_0=10^{-5}$.}
	\label{fig:I_SN}
\end{figure}
%\newpage
%%%%%%%%%%%%%%%%%%%%%
Since Eq.~(\ref{eqn:g_SN}) has the same structure as for a bulk superconductor,  Eq.~(\ref{eqn:g_S}), we expect the presence of a minigap in the quasiparticle spectrum \cite{Belzig1996,Scheer2001}. This feature is clearly seen in Fig.~\ref{fig:SN_DOS} that shows the density of states (DOS) per unit volume, $N(\epsilon)=N_0\Re{g}_N^R(\epsilon)$ for the temperature $T/T_c=0.01$ and several values of $\Gamma/\Delta_0$, with a finite Dynes broadening parameter $\eta/\Delta_0=10^{-3}$. The minigap is narrower than the bulk gap $\Delta$ and for small $\Gamma\ll\Delta_0$ it is determined by $\Gamma$ (see the black line in Fig.~\ref{fig:SN_DOS} that corresponds to $\Gamma/\Delta_0=0.1$). 

$\mathcal{I}(\omega)$ from Eq.~(\ref{eqn:I_S}) is plotted in Fig.~\ref{fig:I_SN}(a) for  the temperature $T/T_c=0.01$ and several values of~$\Gamma$. Similarly to the bulk superconductor case, it has a gap determined by the minigap in the island DOS, strongly dependent on $\Gamma$ [see Fig.~\ref{fig:I_SN}(a)]. For $\Gamma\ll \Delta_0$, the gap is approximately $2\Gamma$ [the black line in Fig.~\ref{fig:I_SN}(a)]. Since we are at low temperature, the gap is empty. The same function for various temperatures and the  $\Gamma/\Delta_0=0.2$ is shown in Fig.~\ref{fig:I_SN}(b). Again like in the bulk case, the gap is getting filled towards the higher temperatures finally achieving the normal state at $T=T_c$. 

 %%%%%%%%%%%%%%%%%%%%%%%%%%%%%%%%%%%%%%%%%%
 \begin{figure}[b]
 	\centering
 	\includegraphics[width=8cm]{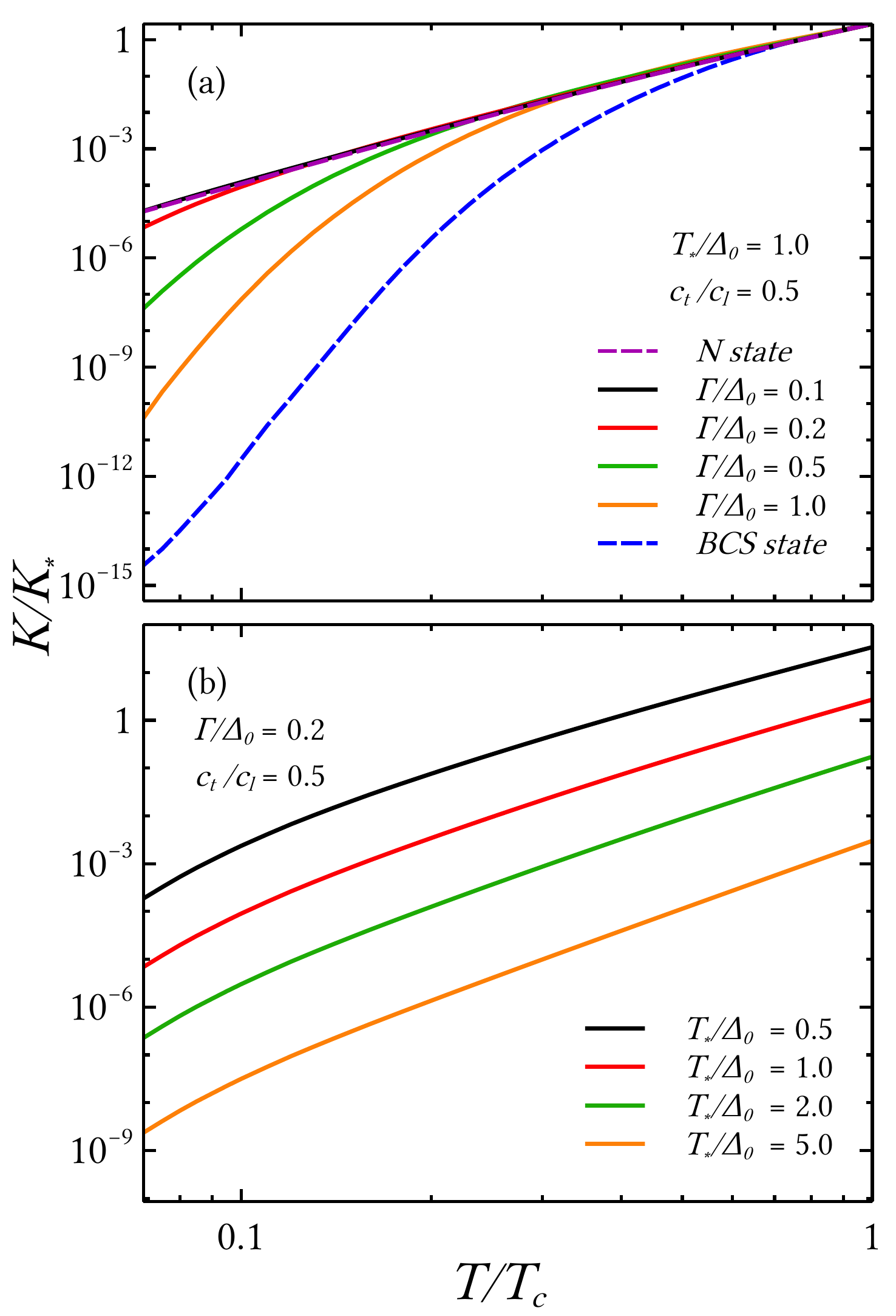}
 	\caption{(Color online) Electron-phonon thermal conductance per unit volume as a function of temperature $T$ in a thin normal metal coupled to a massive superconducting lead for $c_t/c_l = 0.5$ and (a)~different values of $\Gamma/\Delta_0$ and $T_*/\Delta_0=1.0$, (b)~different values of $T_*/\Delta_0$ and $\Gamma/\Delta_0=0.2$. The dashed violet and blue lines correspond to the cases of a bulk normal metal and a bulk superconductor, respectively, at $T_*/\Delta_0=1.0$ and 5.0 on panels (a) and (b), respectively.}
 	\label{fig:G_SN}
 \end{figure}
%%%%%%%%%%%%%%%%%%%%%%%%%%%%%%%%%%%%%%%%%%

Plugging this $\mathcal{I}(\omega)$ into Eq.~(\ref{eqn:G_S}), one arrives at the thermal conductance $K(T)$ per unit volume, shown in Fig.~\ref{fig:G_SN}(a) for various values of $\Gamma/\Delta_0$,  $T_*/\Delta_0=1$, and $c_t/c_l=0.5$. All curves lie between those for the bulk superconductor (the dashed blue line) and the normal state (the dashed violet line) and $K(T)$ is suppressed towards larger~$\Gamma$. This clearly follows from the fact that the minigap in the DOS grows with~$\Gamma$, always remaining smaller then $\Delta_0$ (Fig.~\ref{fig:SN_DOS}). With increasing temperatures all curves tend towards the normal state which is recovered at $T=T_c$. For small~$\Gamma$, e.g. $\Gamma/\Delta_0=0.1$, the minigap is very narrow and this case shows a similar behavior like the normal metal even at low temperatures, $T/T_c\sim 0.07$ [the black line in Fig.~\ref{fig:G_SN}(b)]. On the other hand, for $\Gamma/\Delta_0=1.0$ the minigap is quite large (the blue line in Fig.~\ref{fig:SN_DOS}), and the behavior of the proximitized metal in this case is similar to the bulk superconductor [the orange line in Fig.~\ref{fig:G_SN}(a)].

To see the role of $T^*/\Delta_0$, we plot $K(T)$ in Fig.~\ref{fig:G_SN}(b) for various values of the $\alpha$ parameter and $\Gamma/\Delta_0=0.2$, $c_t/c_l=0.5$. Similarly to the bulk superconductor, increase of $T^*/\Delta_0$ does not change the shape of the curves, but just shifts them downwards.

%\newpage
\subsection{Thin SN bilayer}
\begin{figure}[h]
	\centering
	\includegraphics[width=\columnwidth]{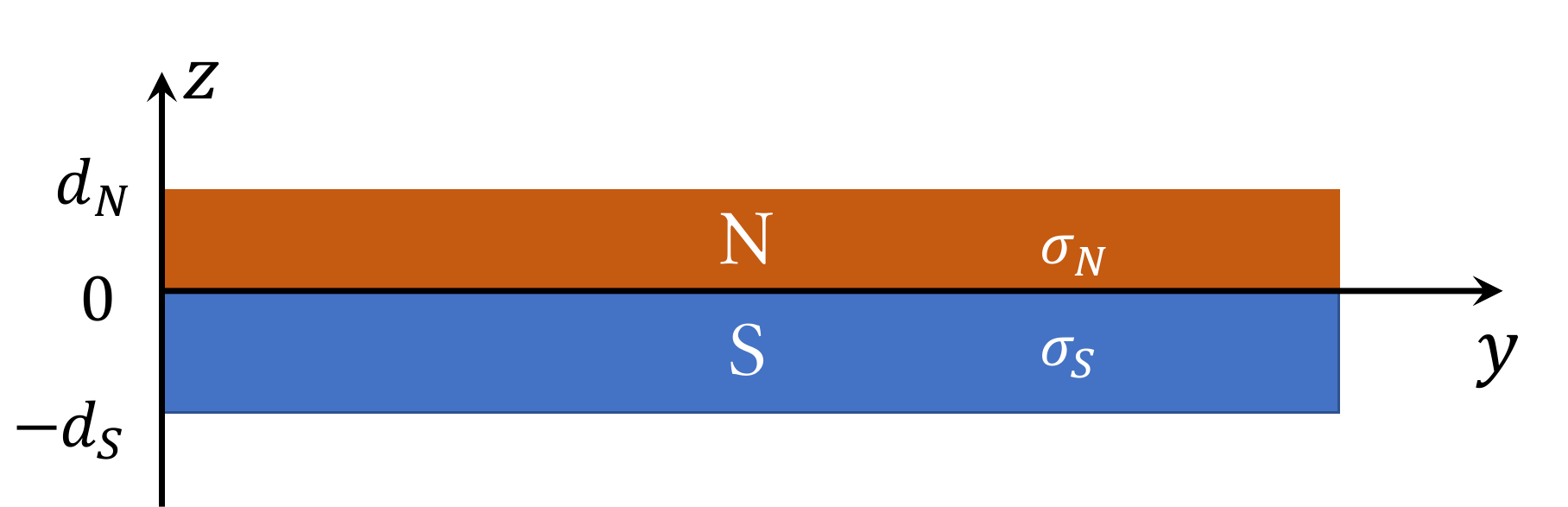}
	\caption{(Color online) A schema of a thin $S\N$ bilayer consisting of a normal metal (orange) of a thickness $d_N$ coupled to a superconductor (blue) of a thickness $d_S$. The nonideal $\S\N$ interface is characterized by the electric conductance $\mathcal{G}$ per unit area, whereas $\sigma_{N/S}$ denotes the normal state conductivity of the $\N/\S$ layer material.}
	\label{fig:SN_bilayer}
\end{figure}

Finally, let us consider a thin $\S\N$ bilayer in the dirty limit $\Delta_0\tau\ll1$, shown schematically in Fig.~\ref{fig:SN_bilayer} and described in the corresponding caption. 
The difference between this geometry and the structure studied in the previous subsection is twofold: (i)~both the normal metal and the superconductor are thin, and (ii)~the contact is not considered in the tunneling limit, i.~e. the interface conductance per unit area~$\mathcal{G}$ is a measure of imperfection of the $\S\N$ interface which would be perfectly transparent in the ideal case with $\mathcal{G}\to\infty$.
Both these features lead to the inverse proximity effect in the superconductor that has now to be treated on equal footing with the proximity effect in the normal layer.

%%%%%%%%%%%%%%%%%%%%%%%%%%%%
%%%%%%%%%%%%%%%%%%%%%%%%%%

The Green's functions for this system were found in Ref.~\cite{Fominov2001}. Assuming the system to be homogeneous in the plane ($x,y$~dimensions), we arrive at an one-dimensional problem along the transverse ($z$) direction.
Parametrizing the Green's function $\hat{g}$ by the proximity angle~$\theta$ such that $\hat{g}=\hat\tau_3\cos\theta+\hat\tau_1\sin\theta$, we can write the Usadel equation in each of the two materials as~\cite{Belzig1999}
\begin{equation}
\label{eqn:Usadel}
\frac{D}{2}\frac{d^2\theta}{dz^2} =-i\epsilon\sin\theta-\Delta\cos\theta,
\end{equation}
where $D=v_F\ell/3$ is the diffusion coefficient of the corresponding material, and $\Delta$ is superconducting gap that is nonzero only for $z<0$. In principle $ \Delta $ has to be determined selfconsistently for a given geometry, which we neglect here for simplicity.
Eq.~(\ref{eqn:Usadel}) should be supplemented by the boundary conditions. At the $\S\N$ interface, $z=0$, we have~\cite{KuprianovLukichev1988}
\begin{equation}\label{eqn:KuprianovLukichev}
\sigma_S\left.\frac{d\theta}{dz}\right|_{z=0^-} = 
\sigma_N\left.\frac{d\theta}{dz}\right|_{z=0^+} = 
\mathcal{G}\sin[\theta(0^+)-\theta(0^-)],
\end{equation}
where $\sigma_{N/S}=2e^2N_{0,N/S}D_{N/S}$ is the normal state conductivity of the normal metal/superconductor. At the free surfaces, $z=-d_S$, $z=d_N$ (Fig.~\ref{fig:SN_bilayer}), there is no current flow and the boundary conditions are simply $(d\theta/dz)|_{z=-d_S} = (d\theta/dz)|_{z=d_N} = 0$.

%%%%%%%%%%%%%%%%%%%%%%%%%%
\begin{figure}[h]
	\includegraphics[width=\columnwidth]{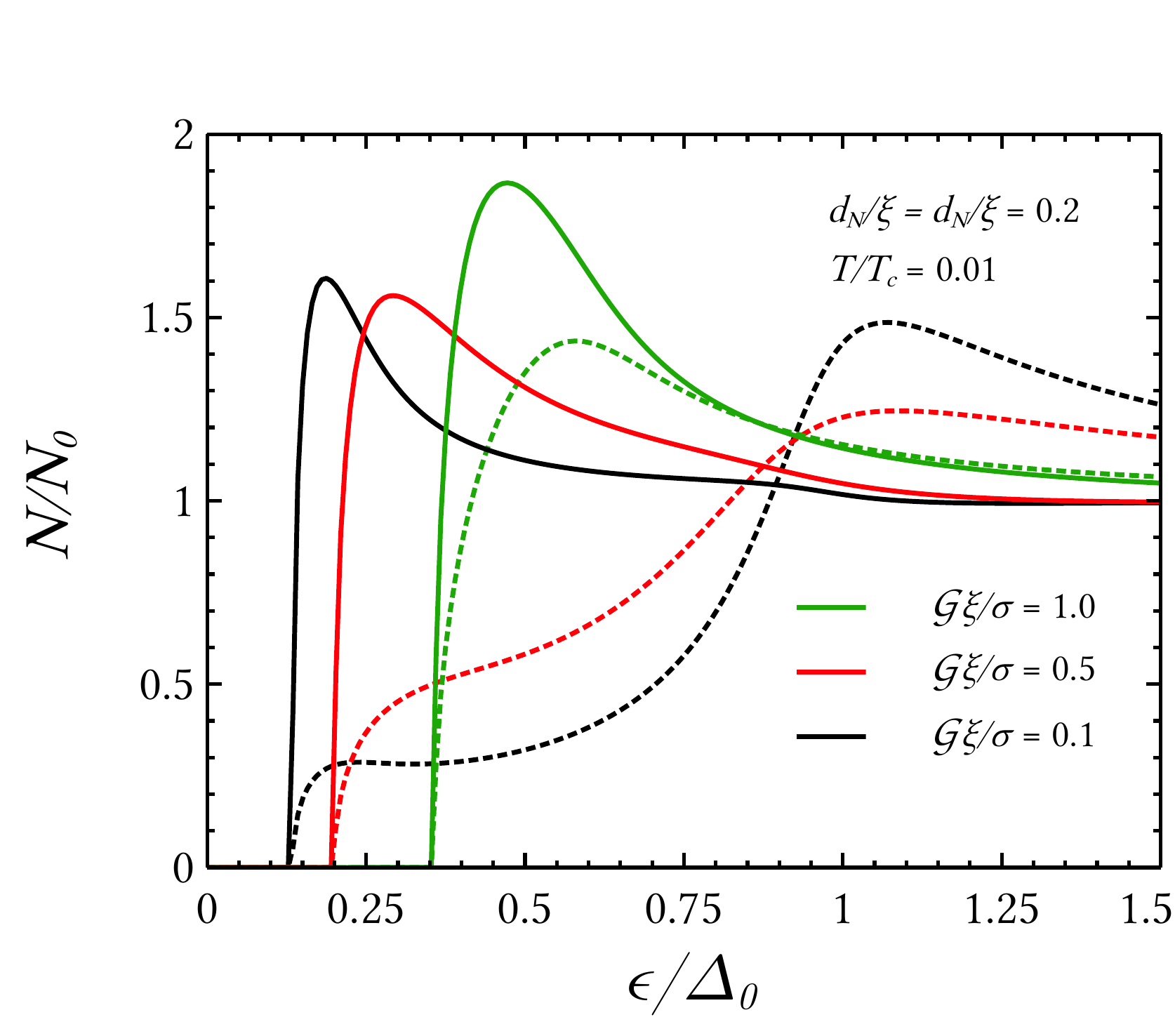}
	\caption{(Color online) Local DOS in the normal (solid lines) and superconducting (dotted lines) parts of a thin SN bilayer for various $\mathcal{G}$, the equal thicknesses of the layers $d_N = d_S=0.2\,\xi$, and the temperature $T/T_c=0.01$.}
	\label{fig:SN_bilayer_DOS}
\end{figure}
%%%%%%%%%%%%%%%%%%%%%%%%%%

Assuming the system to be thinner than the superconducting coherence length, $d_S+d_N \ll \xi\equiv\sqrt{D/\Delta}$, we can seek the solution in the form
\begin{subequations}\label{eqn:thetaexpansion}\begin{align}
&\theta(z<0)=\theta_S+\frac{\theta_S''}2\,(z+d_S)^2+\ldots,\\
&\theta(z>0)=\theta_N+\frac{\theta_N''}2\,(z-d_N)^2+\ldots,
\label{eqn:thetaexpansionN}
\end{align}\end{subequations}
where the second term is small compared to the main one by a factor $\sim{d}_{S,N}^2/\xi^2$, and subsequent terms are even smaller.
Then boundary conditions~(\ref{eqn:KuprianovLukichev}) lead to the following system of nonlinear equations:
%\begin{eqnarray}
%\label{eqn:theta_SN}
%id_N\sigma_N\epsilon\sin(\theta_N) =\ && -id_S\sigma_S\left[\epsilon\sin(\theta_S) + \Delta\cos(\theta_S)\right]=\nonumber
%\\
%=\ &&\frac{gD}{2}\sin(\theta_N-\theta_S),
%\end{eqnarray}
\begin{subequations}\label{eqn:theta_SN}
\begin{align}
&\frac{\mathcal{G}}{4e^2N_{0S}d_S}\,\sin(\theta_S-\theta_N)
=i\epsilon\sin\theta_S+\Delta\cos\theta_S,\quad\label{eqn:Usadel0DS}\\
&\frac{\mathcal{G}}{4e^2N_{0N}d_N}\,\sin(\theta_S-\theta_N)
=-i\epsilon\sin\theta_N.\label{eqn:Usadel0DN}
\end{align}\end{subequations}
The retarded and advanced solutions are obtained by shifting $\epsilon\to\epsilon\pm{i}\eta$. Note that Eq.~(\ref{eqn:Usadel0DN}) has exactly the same form as Eq.~(\ref{eqn:Usadel0D}), with $\Gamma$~given by the coefficient on the left-hand side of Eq.~(\ref{eqn:Usadel0DN}). When the coefficient on the left-hand side of  Eq.~(\ref{eqn:Usadel0DS}) is small compared to $\Delta$, that is $(\mathcal{G}\xi/\sigma_S)(\xi/d_S)\ll1$, then $\theta_S$ is close to its bulk value, and we recover the results of the previous subsection. When $\Gamma\ll\Delta$, the relevant energy scale in the normal metal is $\epsilon\sim\Gamma$, so the length scale controlling the expansion in Eq.~(\ref{eqn:thetaexpansionN}) is $\sqrt{D/\Gamma}$, and the condition $d_N\ll\sqrt{D/\Gamma}$ is equivalent to $\Gamma\ll{E}_{\mathrm{Th}}=D/d_N^2$. In the opposite limit of a thick superconductor, $d_S\gg\xi$, the correction to $\theta_S$ is small and the results of the previous subsection are recovered when $\mathcal{G}\xi/\sigma_S\ll1$.

The local DOS per unit volume can be obtained as $N(\epsilon,z)=N_0\Re\cos\theta^R_N(\epsilon,z)$ [in fact, the $z$ dependence is weak in the regime of the expansion (\ref{eqn:thetaexpansion})].
We take $D_N=D_S=D$, $\sigma_N=\sigma_S=\sigma$, $T=0.01\,T_c$, $d_N = d_S=0.2\,\xi$, 
and plot in Fig.~\ref{fig:SN_bilayer_DOS} $N(\epsilon,d_N)$ (solid lines) and $N(\epsilon,-d_S)$ (dotted lines) as a function of energy for various $\mathcal{G}$.
The main feature as in all gapped systems is the minigap which is smaller than $\Delta_0$ and shrinking as $\mathcal{G}$ decreases. The spectrum in both $\N$ and $\S$ layer is smeared, due the inverse proximity effect, taken into account here and neglected in the previous subsection. More details on these results can be found in Ref.~\cite{Fominov2001}. 

 %%%%%%%%%%%%%%%%%%%%%%%%
 \begin{figure}[t]
 	\centering
 	\includegraphics[width=8cm]{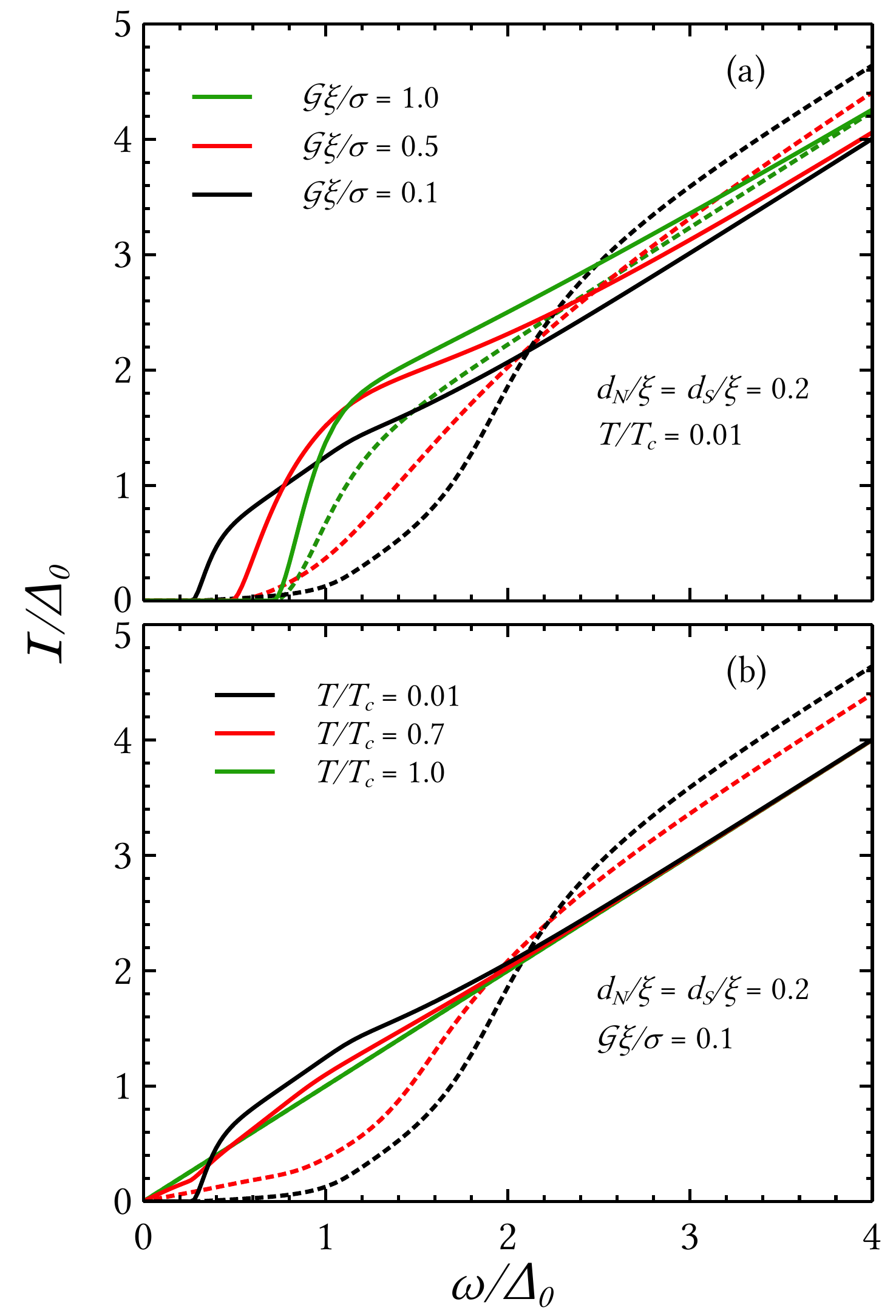}
 	\caption{(Color online)  $\mathcal{I}(d_N,\omega)$ (solid lines) and $\mathcal{I}(-d_S,\omega)$ (dotted lines)  (a)~for different  $\mathcal{G}$ and $T/T_c=0.01$, (b)~for various temperatures and $\mathcal{G}\xi/\sigma = 0.1$. Other parameters are the same as in Fig.~\ref{fig:SN_bilayer_DOS}.}
 	\label{fig:I_SN_bilayer}
 \end{figure}
\begin{figure*}[t]
	\centering
	\includegraphics[width=\linewidth]{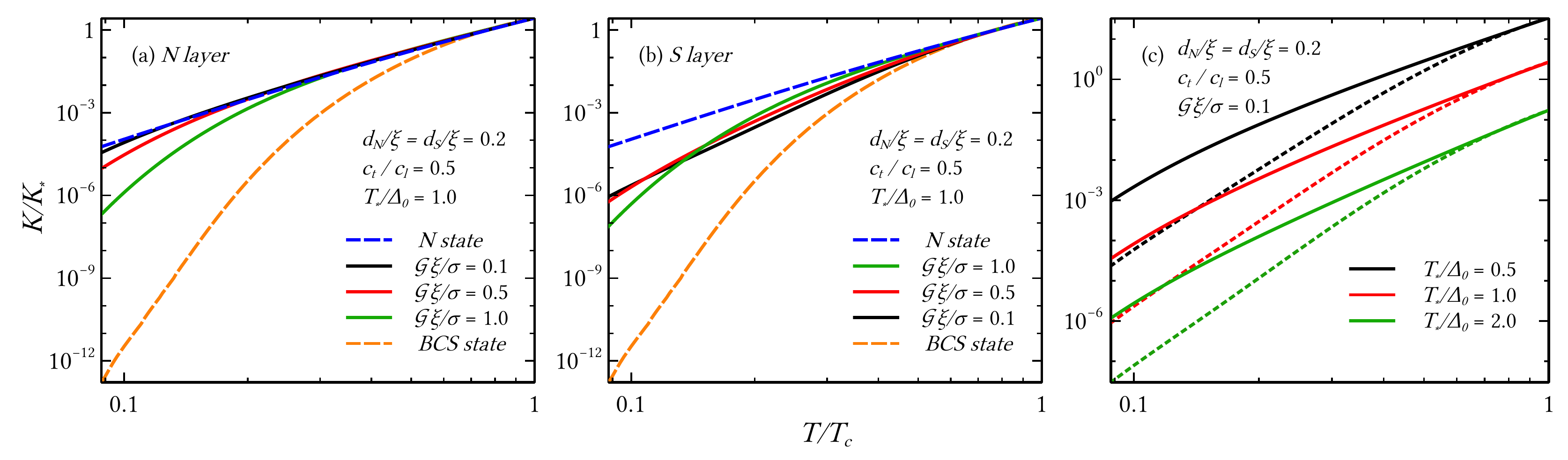}
	\caption{(Color online) Electron-phonon thermal conductance per unit area (a)~on the normal and (b)~superconducting side of a thin $\S\N$ bilayer as a function of temperature $T$ for various $\mathcal{G}$, $c_t/c_l=0.5$, and $T_*/\Delta_0=1.0$. The dashed blue and orange lines correspond to the cases of a bulk normal metal and a bulk superconductor. Panel (c) shows $K(T)$ per unit area on the normal (solid lines) and the superconducting side (dotted lines) of a thin $\S\N$ bilayer for various values of $T_*/\Delta_0$, $c_t/c_l=0.5$, and $\mathcal{G}\xi/\sigma = 0.1$. Other parameters are the same as in Fig.~\ref{fig:SN_bilayer_DOS}.} 
	\label{fig:G_SN_bilayer}
\end{figure*}

 %%%%%%%%%%%%%%%%%%%%%%%%%%%%%%%%%%%%%%%%%
Having found the Green's functions, we calculate $\mathcal{I}(z,\omega)$ from Eq.~(\ref{eqn:I_S}). We plot $\mathcal{I}(d_N,\omega)$ (solid lines) and $\mathcal{I}(-d_S,\omega)$ (dotted lines) for different transparencies [controlled by $\mathcal{G}$] of the $\S\N$ interface [Fig.~\ref{fig:I_SN_bilayer}(a)] and different temperatures [Fig.~\ref{fig:I_SN_bilayer}(b)], other parameters being the same as in Fig.~\ref{fig:SN_bilayer_DOS}. As before, $\mathcal{I}(d_N,\omega)$ exhibits a gap that strongly depends on $\mathcal{G}$. The edge of the gap is not sharp due to the smeared spectrum previously shown in Fig.~\ref{fig:SN_bilayer_DOS}. Increasing temperature leads to shrinking and filling of the gap until $T=T_c$, where we arrive at the normal state in both layers.
%%%%%%%%%%%%%%%%%%%%%%%%%%%%%%%%%%%%%%%%%%
Plugging $\mathcal{I}(z,\omega)$ into Eq.~(\ref{eqn:G_S}), we obtain the electron-phonon thermal conductance $K(T)$ in the normal metal and the superconductor \textit{per unit area} of the structure. In Fig.~\ref{fig:G_SN_bilayer}(a) we plot  $K(T)$ on the normal side for various transparencies of the $\S\N$ interface for $c_t/c_l=0.5$, $T_*/\Delta_0=1.0$ and other parameters as in Fig~\ref{fig:SN_bilayer_DOS}. The minigap suppresses the cooling power at low temperatures. With increasing $\mathcal{G}$ the effect is stronger since the minigap is getting larger. In Fig.~\ref{fig:G_SN_bilayer}(b) we present $K(T)$ on the superconducting side for the same parameters as in Fig.~\ref{fig:G_SN_bilayer}(a). As in all superconducting structures, at $T=T_c$, all curves converge to the normal state one. Since the minigap is always smaller than $\Delta$, the low-temperature suppression of $K(T)$ is weaker than in the bulk superconductor case [the dashed orange line in Fig.~\ref{fig:G_SN_bilayer}(a,b)] but, depending on $\mathcal{G}$, much stronger than in the normal case [the blue line in Fig.~\ref{fig:G_SN_bilayer}(a,b)].
Fig.~\ref{fig:G_SN_bilayer}(c) shows $K(T)$ on the normal (solid lines) and the superconducting side (dotted lines) of a thin $\S\N$ bilayer for various values of $T_*/\Delta_0$ and the conductance of the $\S\N$ interface $\mathcal{G}\xi/\sigma=0.1$. As in the previous subsections increasing $T_*/\Delta_0$ just shifts the curves downwards. One notes that the effect is stronger in the $S$ region due to the inverse proximity effect visible in Fig.~\ref{fig:SN_bilayer_DOS}.

%%%%%%%%%%%%%%%%%%%%%%%%%%%%%%%%%%
%\clearpage
\section{Conclusions}
\label{sec:conclusion}

We have studied electron cooling by phonons in superconducting proximity structures. Using the quasiclassical approximation and perturbation theory in electron-phonon coupling, we obtained a rather general formula for the cooling power and the thermal conductance, Eq.~(\ref{eqn:heat_current}), that is applicable to an arbitrary electronic system, even nonequilibrium. We focused on situations when electrons and phonons are in equilibrium among themselves, but have different temperatures. In the simple cases of a bulk normal metal and a bulk BCS superconductor we recovered the previously known results. 

Subsequently, we illustrated our theory on two simple geometries of a superconductor-normal metal contact. Due to the presence of a proximity minigap, these heterostructures exhibit a strong suppression of the cooling power at low temperatures which makes them suitable candidates for making quantum thermal detectors. Our theory can serve as a tool for optimizing the structure in order to improve the detector sensitivity, which could serve as a benchmark for future experiments. 

\acknowledgements
We thank Jukka Pekola and Bayan Karimi for numerous useful discussions. D.N. thanks Universit\'e Grenoble Alpes and CNRS for hospitality during his visit.
This work was funded through the European Union’s Horizon 2020 research and innovation programme under Marie Sklodowska-Curie actions (Grant No. 766025). 
\appendix

\section{Keldysh Green's functions}\label{app:Keldysh}
In the derivation of Eq.~(\ref{eqn:Q_ph_K}) based on the Kubo formula we make use of the contour-ordered Green's function which can be constructed for an arbitrary set of bosonic fields $\varphi_\alpha(t)$ (the index $\alpha$ incorporating the spatial coordinates and all other indices) as
\begin{subequations}\begin{eqnarray}
	&&\breve{D}_{\alpha\beta}(t,t')= \left(\begin{array}{cc} D_{\alpha\beta}(t,t') & D_{\alpha\beta}^{<}(t,t')\\ D_{\alpha\beta}^{>}(t,t') & \tilde{D}_{\alpha\beta}(t,t') \end{array}\right),\\
	&&iD_{\alpha\beta}(t,t') = \langle \mathcal{T} \varphi_{\alpha}(1) \varphi_{\beta}(1')  \rangle,
	\\
	&&iD_{\alpha\beta}^{<}(t,t') = \langle\varphi_{\beta}(1') \varphi_{\alpha}(1)\rangle,
	\\
   	&&iD_{\alpha\beta}^{>}(t,t') = \langle \varphi_{\alpha}(1)\varphi_{\beta}(1')\rangle,
   	\\
   	&&i\tilde{D}_{\alpha\beta}(t,t') = \langle \mathcal{\tilde{T}} \varphi_{\alpha}(1) \varphi_{\beta}(1')  \rangle,
\end{eqnarray}\end{subequations}
where $\mathcal{T}~(\mathcal{\tilde{T}})$ denotes chronological (antichronological) time ordering. These functions are not independent and by performing the Larkin-Ovchinnikov rotation \cite{Larkin1975}, $\check{D}\rightarrow \check{L}\check{\tau}_3 \check{D} \check{L}^\dagger$ with $\check{L}=(1 - i\check{\tau}_2)/\sqrt{2}$ and $\check{\tau}_2$ being the second Pauli matrix, we pass to the so-called Keldysh space obtaining
\begin{equation}
	\check{D}_{\alpha\beta}(t,t') = \left(\begin{array}{cc} D^R_{\alpha\beta}(t,t') & D^K_{\alpha\beta}(t,t')\\ 0 & D^A_{\alpha\beta}(t,t') \end{array}\right).
\end{equation}
The newly introduced functions satisfy the following relations [$\theta(t)$ being the Heaviside step function]:
\begin{eqnarray}
D^R_{\alpha\beta}(t,t') &&= \theta(t-t')\left[D^>_{\alpha\beta}(t,t') - D^<_{\alpha\beta}(t,t')\right],
\\
D^A_{\alpha\beta}(t,t') &&= \theta(t'-t)\left[D^<_{\alpha\beta}(t,t') - D^>_{\alpha\beta}(t,t')\right],
\\
D^K_{\alpha\beta}(t,t') &&= D^>_{\alpha\beta}(t,t') + D^<_{\alpha\beta}(t,t'),
\end{eqnarray}
which are used to derive Eq.~(\ref{eqn:Q_ph_K}) from Eq.~(\ref{eqn:Kubo}).

\section{Calculation of the $\mathcal{Y}_\lambda(q\ell)$ factors}
\label{appendix:qL_factors}
The factors $\mathcal{Y}_\lambda(q\ell)$ coming from angular averages that are given by
\begin{widetext}
\begin{equation}
\mathcal{Y}_\lambda(q\ell)=
\left\langle\frac{\Phi_\lambda(\vec{n})}{1+\ell^2(\vec{q}\vec{n})^2}\right\rangle_\vec{n}^2
\left(1-\left\langle\frac{1}{1+\ell^2(\vec{q}\vec{n})^2}\right\rangle_\vec{n}\right)^{-1}+\left\langle\frac{\Phi_\lambda^2(\vec{n})}{1+\ell^2(\vec{q}\vec{n})^2}\right\rangle_\vec{n},
\end{equation}
\end{widetext}
where $\langle \dots \rangle_\vec{n}$ denotes averaging over the directions of the Fermi velocity and $\Phi_\lambda(\vec{n})$ are given in Eq.~(\ref{eqn:Phi_lambda}).
Evaluation of the angular averages gives
%\begin{widetext}
%\begin{eqnarray}
%&&\left\langle\frac{1}{1+\ell^2(\vec{q}\vec{n})^2}\right\rangle_\vec{n}
%=\frac{\arctan{q}\ell}{q\ell},\\
%&&\left\langle\frac{(\vec{q}\vec{n})^2/q^2-1/3}{1+\ell^2(\vec{q}\vec{n})^2}\right\rangle_\vec{n}=
%\frac{q\ell-(1+q^2\ell^2/3)\arctan{q}\ell}{q^3\ell^3},\\
%&&\left\langle\frac{[(\vec{q}\vec{n})^2/q^2-1/3]^2}%
%{1+\ell^2(\vec{q}\vec{n})^2}\right\rangle_\vec{n}=
%\frac{1+q^2\ell^2/3}{(q\ell)^5}\left[-q\ell+(1+q^2\ell^2/3)\arctan{q}\ell\right],\\
%&&\mathcal{Y}_l(q\ell)=-\frac{q\ell-(1+q^2\ell^2/3)\arctan{q}\ell}{3q^2\ell^2(q\ell-\arctan{q}\ell)},\\
%&&\left\langle\frac{(\vec{e}_{t\kappa}\vec{n})(\vec{q}\vec{n})/q}{1+\ell^2(\vec{q}\vec{n})^2}\right\rangle_\vec{n}=0,\\
%&&\left\langle\frac{(\vec{e}_{t\kappa}\vec{n})^2(\vec{q}\vec{n})^2/q^2}%
%{1+\ell^2(\vec{q}\vec{n})^2}\right\rangle_\vec{n}=
%\frac{q\ell(1+2q^2\ell^2/3)-(1+q^2\ell^2)\arctan{q}\ell}{2q^5\ell^5}
%=\mathcal{Y}_{t1,t2}(q\ell).
%\end{eqnarray}
%\end{widetext}
\begin{align*}
&\left\langle\frac{1}{1+\ell^2(\vec{q}\vec{n})^2}\right\rangle_\vec{n}
=\frac{\arctan{q}\ell}{q\ell},\\
&\left\langle\frac{(\vec{q}\vec{n})^2/q^2-1/3}{1+\ell^2(\vec{q}\vec{n})^2}\right\rangle_\vec{n}=
\frac{q\ell-(1+q^2\ell^2/3)\arctan{q}\ell}{q^3\ell^3},
\end{align*}
\begin{align*}
&\left\langle\frac{[(\vec{q}\vec{n})^2/q^2-1/3]^2}%
{1+\ell^2(\vec{q}\vec{n})^2}\right\rangle_\vec{n}=
\frac{(1+q^2\ell^2/3)^2}{(q\ell)^5}\,\arctan{q}\ell-{}\nonumber\\
&\hspace*{4cm}{}-\frac{1+q^2\ell^2/3}{(q\ell)^4},\\
&\left\langle\frac{(\vec{e}_{t\kappa}\vec{n})(\vec{q}\vec{n})/q}{1+\ell^2(\vec{q}\vec{n})^2}\right\rangle_\vec{n}=0,
\end{align*}
\begin{subequations}\begin{align}
&\mathcal{Y}_l(q\ell)=-\frac{q\ell-(1+q^2\ell^2/3)\arctan{q}\ell}{3q^2\ell^2(q\ell-\arctan{q}\ell)},\\
&\mathcal{Y}_{t1,t2}(q\ell)=
\left\langle\frac{(\vec{e}_{t\kappa}\vec{n})^2(\vec{q}\vec{n})^2/q^2}%
{1+\ell^2(\vec{q}\vec{n})^2}\right\rangle_\vec{n}={}\nonumber\\
&\hspace*{1.5cm}{}=\frac{q\ell(1+2q^2\ell^2/3)-(1+q^2\ell^2)\arctan{q}\ell}{2q^5\ell^5}.
\end{align}\end{subequations}
Note that these integrals are the same for a normal metal and a superconductor, and $\mathcal{Y}_l(q\ell)$ are in agreement with Ref.~\cite{Schmid1973}. The asymptotic behaviour is
\begin{subequations}\begin{eqnarray}
\label{eqn:Ql0}
&&\mathcal{Y}_l(q\ell\to0)=\frac{4}{45}+O(q^2\ell^2),\\
\label{eqn:Qlinf}
&&\mathcal{Y}_l(q\ell\to\infty)=\frac{\pi}{18\,q\ell}+O(q^{-2}\ell^{-2}),
\end{eqnarray}
\begin{eqnarray}
\label{eqn:Qt0}
&&\mathcal{Y}_{t1,t2}(q\ell\to0)=\frac{1}{15}+O(q^2\ell^2),\\
\label{eqn:Qtinf}
&&\mathcal{Y}_{t1,t2}(q\ell\to\infty)=\frac{1}{3\,q^2\ell^2}+O(q^{-3}\ell^{-3}).
\end{eqnarray}\end{subequations}

\end{document}